\documentclass[reprint,amsmath,amssymb,aps,prb, superscriptaddress,twocolumn]{revtex4-1}
\usepackage{graphicx}
\usepackage[colorlinks=true,citecolor=blue,linkcolor=blue,urlcolor=blue]{hyperref}

\usepackage[utf8]{inputenc}
\usepackage{amstext}			
\usepackage{amssymb}			
\usepackage{amsmath}			
\usepackage{amsfonts}
\usepackage{graphicx}
\usepackage{subfigure}
\usepackage{wrapfig}
\usepackage{esint}
\usepackage{multirow}
\usepackage[english]{babel}

\DeclareMathOperator{\tr}{Tr}

\newcommand{\lint}{\int\limits}
\newcommand{\lsum}{\sum\limits}
\newcommand{\BH}{\mathcal{H}}

\newcommand{\BN}{\mathcal{N}}
\newcommand{\BQ}{\mathcal{Q}}
\newcommand{\BC}{\mathcal{C}}

\newcommand{\pmark}[1]{{\bf (#1)}}
\newcommand{\tprime}{$^\prime$}

\begin{document}
  \title{Hybrid quantum-classical method for simulating high-temperature dynamics of nuclear spins in solids}
  \date{\today}
  \author{Grigory A. Starkov}
  \email{grigory.starkov@skolkovotech.ru}
  \affiliation{Skolkovo Institute of Science and Technology, 
Skolkovo Innovation Centre, Nobel Street 3, Moscow 143026, Russia}
\affiliation{Lebedev Physical Institute of the Russian Academy of Sciences, Leninsky prospect 53,
Moscow 119991, Russia}
  \author{Boris V. Fine}
  \email{b.fine@skoltech.ru}
   \affiliation{Skolkovo Institute of Science and Technology, 
Skolkovo Innovation Centre, Nobel Street 3, Moscow 143026, Russia}
\affiliation{Institute for Theoretical Physics, University of Heidelberg, Philosophenweg 12, 69120 Heidelberg, Germany}

 \begin{abstract}

First-principles calculations of high-temperature spin dynamics in solids in the context of nuclear magnetic resonance (NMR) is a long-standing problem, whose conclusive solution can significantly advance  the applications of NMR as a diagnostic tool for material properties. In this work, we propose a new hybrid quantum-classical method for computing NMR free induction decay(FID) for spin $1/2$ lattices. The method is based on the simulations of a finite cluster of spins $1/2$ coupled to an environment of interacting classical spins {\it via} a correlation-preserving scheme. Such simulations are shown to lead to accurate FID predictions for one-, two- and three-dimensional lattices with a broad variety of interactions. The accuracy of these predictions can be efficiently estimated by varying the size of quantum clusters used in the simulations. 

\end{abstract}

    \maketitle


\section{Introduction}  
  
Free induction decay (FID) measured by nuclear magnetic resonance (NMR) is, normally, proportional to the infinite-temperature time autocorrelation function of the total nuclear spin polarization of the system \cite{Lowe-57,Abragam-61}.  It depends on the internuclear distances and spin-spin interactions. The Fourier transform of the FID gives NMR absorption lineshape\cite{Bloch-46-1,Lowe-57,Abragam-61}.  First-principles calculation of NMR FID in solids is a non-perturbative problem --- it does not have a small parameter to build a controllable analytic expansion.
  The problem is normally non-integrable at the quantum level \cite{Fine-14} and chaotic at the classical level \cite{deWijn-12,deWijn-13}.
  It belongs to a broader class of problems exhibiting non-Markovian dynamics,
  often accompanied by non-universal observable behaviour.  
  A number of first-principles methods of FID calculations have been proposed in the past \cite{VanVleck-48,Lowe-57,Abragam-61,Tjon-66,Parker-73,Jensen-73,Engelsberg-75,Becker-76,Shakhmuratov-91,Lundin-92,Jensen-95,Fine-97,Zhang-07,Savostyanov-14,Elsayed-15}.
  Quite a few of them produced good approximations for FID in one system, namely, CaF$_2$\cite{Lowe-57,Engelsberg-74}.
  Yet, none of them is widely used at present, because their predictive
  performance for a broader class of systems is either poor or unclear.
  In the present work, we propose a new hybrid quantum-classical method  of simulating high-temperature spin dynamics that meets the challenge of predictive performance in two ways: the method is tested for one-, two- and three-dimensional spin-1/2 lattices with a broad variety of interactions,
  and, simultaneously, it is shown that one can make an efficient uncertainty estimate for the computed quantity. The defining feature of the method is the implementation of the dynamical action of the quantum cluster on the classical environment.

The method of hybrid simulations is likely to be applicable beyond solid-state NMR  to describe, for example, quantum decoherence\cite{Coish-05,Dobrovitski-06,Zhang-07,Liu-07,Stanwix-10} and inelastic magnetic neutron scattering at high temperatures\cite{Balcar-81}. The advantage of developing the method in the context of NMR is the availability of a very accurate experimental testing ground, which is consequence of the fact that nuclear spin dynamics is well isolated from the electronic and phononic environments.
  

  



\section{Model}
 We consider a lattice of spins $1/2$ with translationally invariant Hamiltonian of the general form:
  \begin{equation}
    {\BH} = \lsum_{\alpha,i<j} J_{i,j}^\alpha {S}_i^{\alpha}{S}_j^{\alpha},\qquad \alpha\in\{x,y,z\},
    \label{Ham}
  \end{equation}
  where $S_i^\alpha$ is the operator of spin projection on axis $\alpha$ for the $i$-th lattice site, and $J^{\alpha}_{i,j}$ are the coupling constants. The quantities of our interest are time autocorrelation functions of the total spin polarization ${M}_\alpha = \sum_i{S}_i^\alpha$
  \begin{equation}
   C_\alpha(t) = \left\langle M_\alpha(t)M_\alpha(0)\right\rangle/\left\langle M^2_\alpha \right\rangle ,
   \label{corf1}
  \end{equation}
where $\langle ... \rangle$ denotes the averaging over the infinite temperature equilibrium state. In general, $C_\alpha(t)$ decays on the fastest microscopic timescale of the system characterized by the inverse root-mean-squared value of local fields experienced by each spin:
\begin{equation}
 \tau_c= \left(\lsum_j {J_{ij}^x}^2 \langle {S_{j}^x}^2\rangle+
                                  {J_{ij}^y}^2 \langle {S_{j}^y}^2\rangle+
                                  {J_{ij}^z}^2 \langle {S_{j}^z}^2\rangle\right)^{-1/2}.
\label{tauc}
\end{equation}
Direct numerical calculation of $C_\alpha(t)$ in the thermodynamic limit is not feasible due to the exponentially large Hilbert spaces involved.





\section{Hybrid method.}  

\begin{figure}\setlength{\unitlength}{0.1in}
\begin{picture}(34,34)
\put(0, 0) {\includegraphics[width=3.4in]{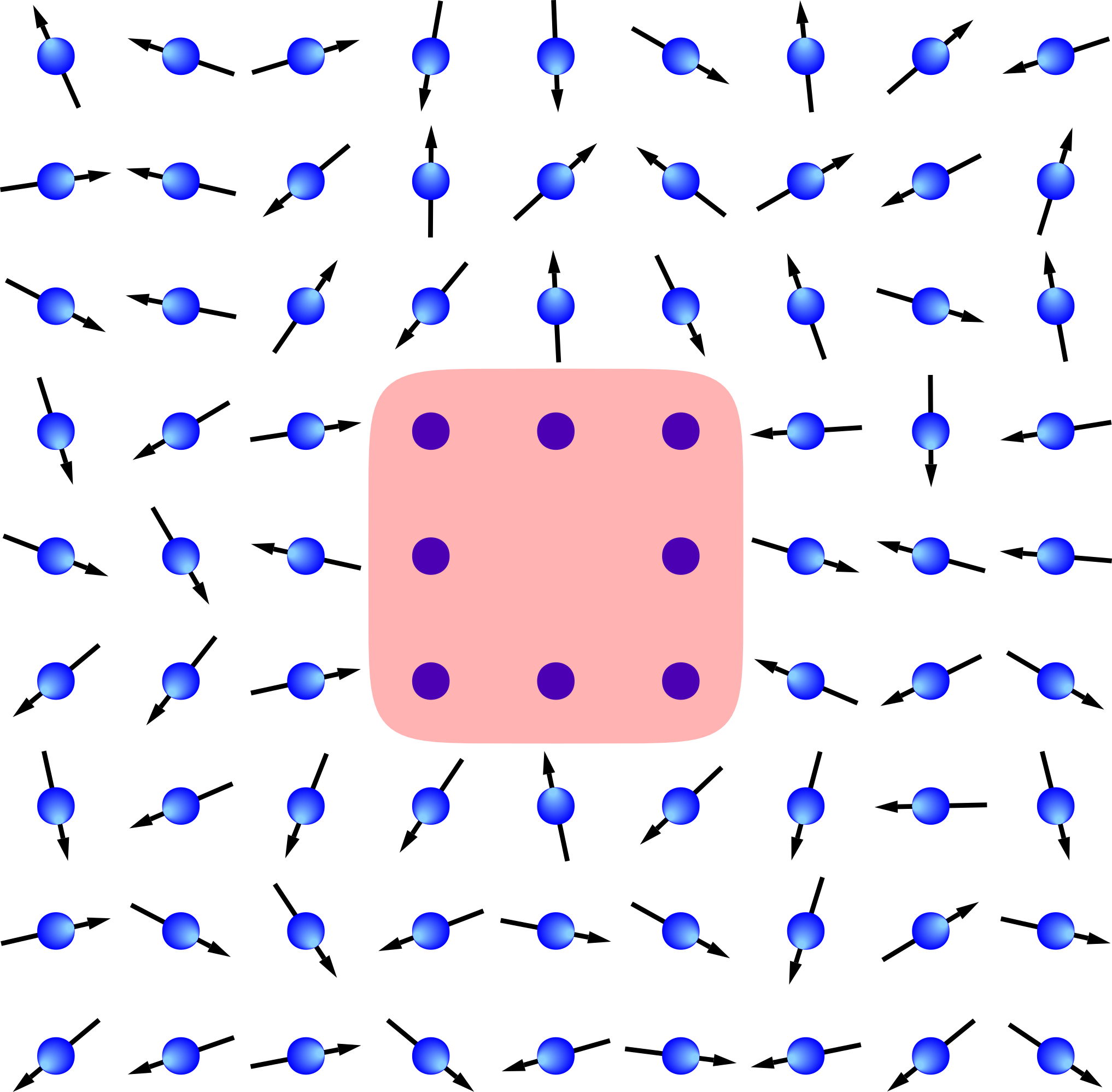}}
 \put(15,17){\raisebox{-0.5\height}{\mbox{\Huge $\mathbf{|\psi\rangle}$}}}
\end{picture}
 \caption{Sketch of a hybrid lattice: a cluster of spins 1/2   surrounded by an environment of classical spins. The quantum cluster is described by a wave function $|\psi\rangle$. Classical spins are represented by three-dimensional vectors.}
 \label{hlattice}
\end{figure}

We replace the above quantum lattice with a hybrid lattice that contains a set of lattice sites $\BQ$ occupied by a cluster of quantum spins 1/2  and a set of sites $\BC$ occupied by classical spins (see Fig.~\ref{hlattice}). The quantum cluster is described by a wave function $|\psi\rangle$, while the classical spins are described by a set of vectors $\left\{ {\bf s}_m \right \}$.
The time evolution of $|\psi\rangle$ is computed quantum mechanically by direct integration of the Schr\"odinger equation, and, simultaneously, the dynamics of the classical spin vectors $\left\{ {\bf s}_m \right \}$ is obtained by the integration of the classical equations of motion (see Appendices~\ref{classical},\ref{quantum},\ref{numerical}).

The challenge in defining the dynamics of such a hybrid system is to reproduce dynamical correlations of the original fully quantum lattice as closely as possible. An important aspect of these correlations is the retarded action of each spin on itself and remote spins via interacting neighbors. In order to induce such correlations across the quantum-classical border, we introduce effective local fields exerted by the two parts on each other.
The local fields exerted by the classical environment on quantum spins are to have the standard form used in purely classical simulations.
In order to define the reverse action of the quantum spins on the classical neighbors, one can try to take the expressions for the classical local fields  and replace there classical spin projections $s^{\alpha}_m$ with the expectation values of quantum spin operators $\langle\psi| S_m^\alpha|\psi\rangle$. However, the problem with such an approach is that, for a typical pure state describing a cluster of $N_\BQ$ spins 1/2, the expectation values $\langle\psi| S_m^\alpha|\psi\rangle $ are exponentially small\cite{Gemmer-04,Pastawski-08,Elsayed-13}: they are  of the order $1/\sqrt{N}$, where $N = 2^{N_{\BQ}}$ is cluster's Hilbert space dimension (see Supplementary Information).    Therefore, such a naive approach would lead to a negligible action of quantum spins on the classical ones, thereby failing to induce qualitatively important correlations across the quantum-classical border.
Instead, we propose to use the quantum expectation values scaled up by factor $\sqrt{N}$, whenever they are coupled to or combined with the classical variables. This rescaling is to be justified after we introduce the formalism.

The dynamics of the quantum and classical parts are described by respective Hamiltonians
  \begin{align}
   {\BH}_\BQ & = \lsum_{i<j,\alpha}^{i,j\in\BQ} J^\alpha_{i,j} {S}^\alpha_i {S}^\alpha_i - \lsum_{i\in\BQ} {\bf h}_i^{\BC\BQ} \cdot {\bf S}_i,\label{q_ham}\\
   {\BH}_\BC         & = \lsum_{m<n,\alpha}^{m,n\in\BC} J^\alpha_{m,n} s^\alpha_m s^\alpha_n -
                            \lsum_{m\in\BC} {\bf h}^{\BQ\BC}_m\cdot {\bf s}_m,\label{cl_ham}
 \end{align}
where ${S}^\alpha_i$ are the operators of spins 1/2 as in Eq.(\ref{Ham}), ${\bf s}_m\equiv (s_m^x,s_m^y,s_m^z)$ are vectors of length $\sqrt{S(S+1)} = \sqrt{3}/2$ representing  classical spins, ${\bf h}^{\BC\BQ}_i$ and ${\bf h}^{\BQ\BC}_i$ are the local fields coupling the quantum and the classical parts:
 \begin{align}
   {\bf h}^{\BC\BQ}_i & = -\lsum_{n\in\mathcal{C}}\left(\begin{array}{c}
                                             J_{i,n}^x s_n^x \\
                                             J_{i,n}^y s_n^y \\
                                             J_{i,n}^z s_n^z \\
                                            \end{array}
\right),\label{hfield_cl}\\
   {\bf h}_m^{\BQ\BC} & = -\sqrt{N}\cdot\lsum_{j\in\mathcal{Q}}\left(\begin{array}{c}
                        J_{m,j}^x \langle\psi|{S}_j^x|\psi\rangle \\
                        J_{m,j}^y \langle\psi|{S}_j^y|\psi\rangle \\
                        J_{m,j}^z \langle\psi|{S}_j^z|\psi\rangle \\
         \end{array}
\right).\label{backaction}
  \end{align}
The lattice has periodic boundary conditions.
  
The initial conditions for the simulations include a fully random choice of $|\psi(0)\rangle$ in the Hilbert space of the quantum cluster and random orientations of classical spins.
The hybrid version of the total spin polarization $M_\alpha(t)$ is defined according to the earlier prescription for rescaling quantum expectation values:
  \begin{equation}
  M_\alpha(t) = \sqrt{N}\cdot\langle\psi(t)|\lsum_{i\in\BQ} {S}_i^\alpha|\psi(t)\rangle
  + \lsum_{m\in\BC} s_m^\alpha(t).
  \label{total_spin}
  \end{equation}
  
  The mathematical construction based on Eqs.(\ref{q_ham}, \ref{cl_ham}, \ref{hfield_cl}, \ref{backaction}, \ref{total_spin})
  introduces dynamical correlations across quantum-classical boundary, which, while being approximate, exactly capture two important aspects of the fully quantum dynamics.
  First, the root-mean-squared value of the local field for each spin, quantum or classical, is the same as that for the original quantum lattice.
  Second, if the Hamiltonian of the original quantum lattice conserves the total spin polarization, or one of its projections, this conservation law is also respected by the hybrid dynamics for $M_\alpha$ defined by Eq.(\ref{total_spin}).
  
  Yet the quantum-classical border still disturbs the dynamics of spins within the quantum cluster in comparison with the purely quantum lattice. This distortion is weaker for the spins located further from the border. Therefore, as explained in Appendix~\ref{representations},  we reduce the influence of the border by introducing an auxiliary variable 
  $M_\alpha' = \sqrt{N}\cdot\langle\psi(t)|\lsum_{m\in\BQ'} {S}_m^\alpha|\psi(t)\rangle$, where the subset $\BQ'$ is limited to one or several central spins within the quantum cluster. We then compute the correlation function of interest as
  \begin{equation} 
  C_{\alpha} (t) = \langle M_\alpha(t) M_\alpha'(0) \rangle/\langle {M_\alpha'}^2 \rangle
  \label{Cprime}
  \end{equation}
   by performing averaging over the equilibrium noise of $M_\alpha(t)$ and $M_\alpha'(t)$.

\begin{figure*} \setlength{\unitlength}{0.1in}
\begin{picture}(70,55)

 \put(6, 36.5) {\includegraphics[width=2.8in]{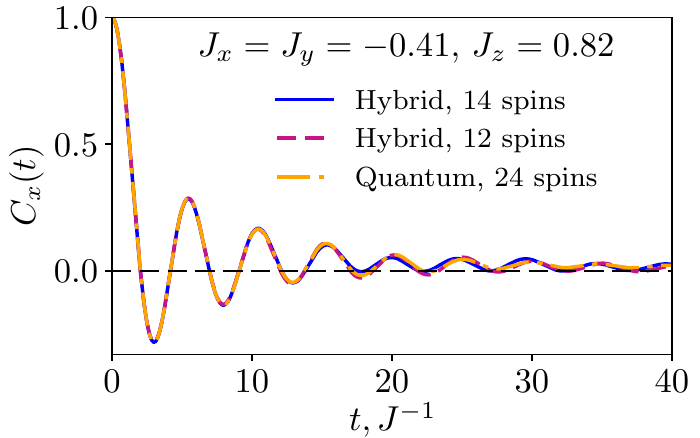}}
 \put(35,36.5) {\includegraphics[width=2.8in]{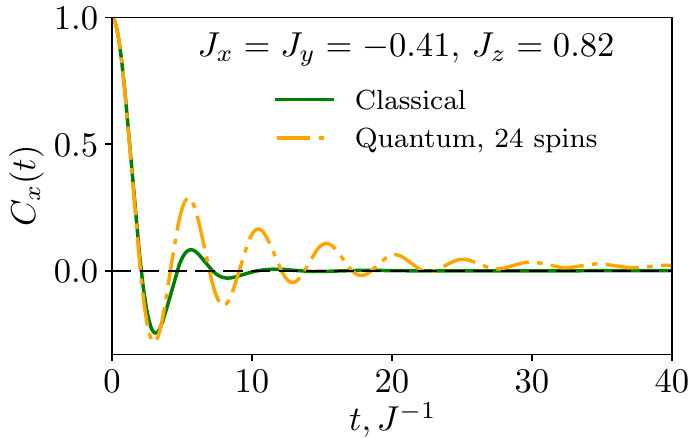}}
 
 \put(6, 18.5) {\includegraphics[width=2.8in]{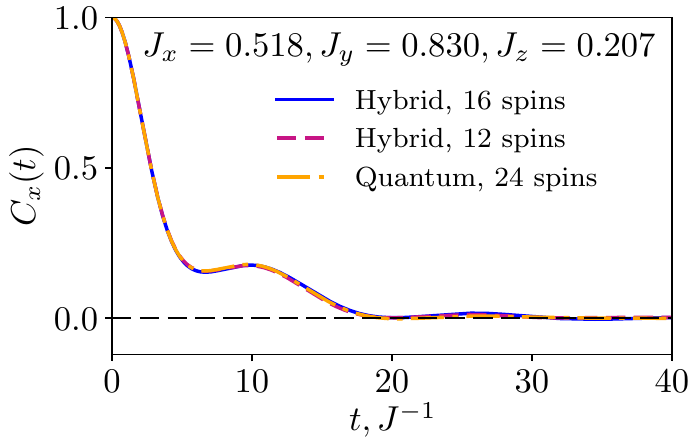}}
 \put(35, 18.5) {\includegraphics[width=2.8in]{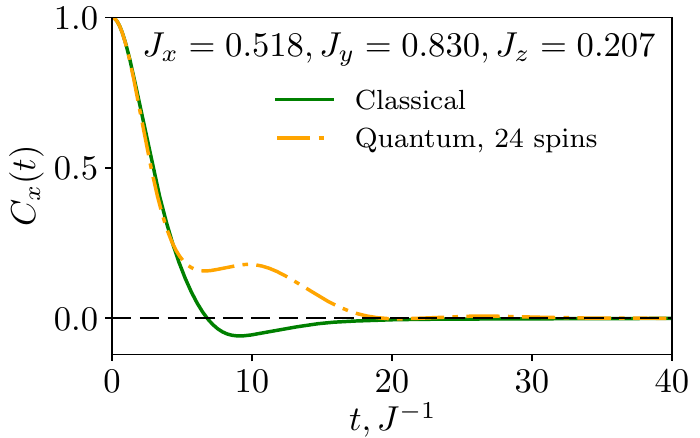}}
 
 \put(6, 0.5) {\includegraphics[width=2.8in]{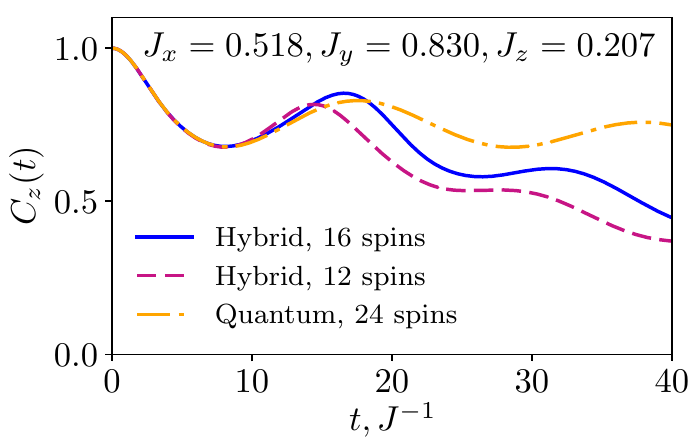}}
 \put(35,0.5) {\includegraphics[width=2.8in]{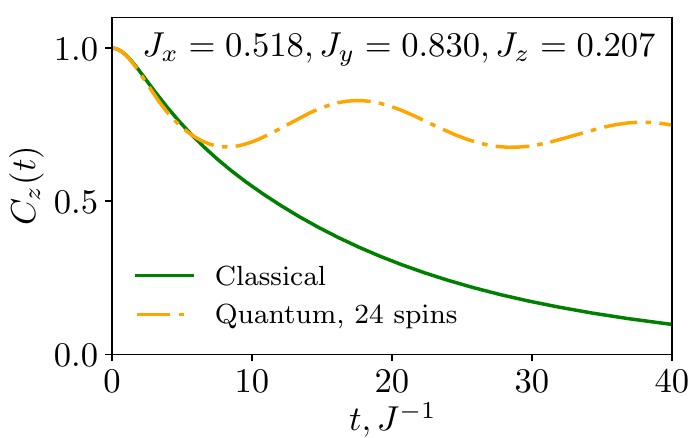}}

 \put(5.5, 54)  {\pmark a}
 \put(34.5, 54) {\pmark {a\tprime}}
 
 \put(5.5, 36)  {\pmark b}
 \put(34.5, 36) {\pmark {b\tprime}}
 
 \put(5.5, 18)  {\pmark c}
 \put(34.5, 18) {\pmark {c\tprime}}
\end{picture}
\caption{Correlation functions $C_{\alpha} (t) $  for one-dimensional periodic chains with nearest neighbours interactions. The interaction constants are indicated above each plot. The left column of plots compares the results of hybrid simulations with the reference plots obtained by direct quantum calculations. The right column does the same for purely classical simulations.
For both hybrid and classical simulations, the full lattice size is $92$. The sizes of quantum clusters in hybrid simulations and in reference quantum calculations are indicated in the plot legends. 
        }

\label{onedim}
\end{figure*}
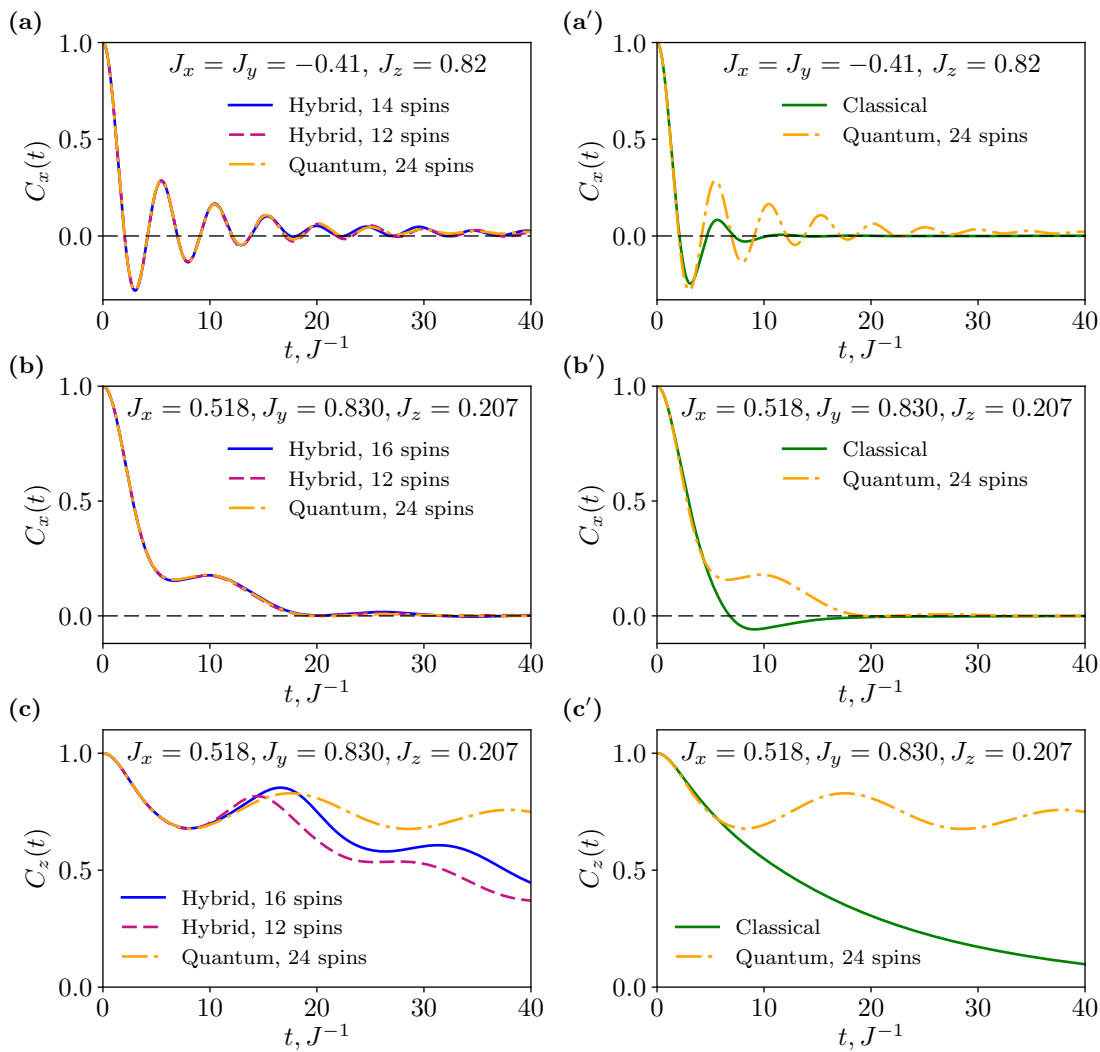

 An important aspect of the hybrid method is that it is possible to make an efficient estimate of the accuracy of its predictions. 
 This estimate is based on the observation that, as the size of the quantum cluster increases, the hybrid calculation must converge to the exact quantum result. Therefore, a discrepancy between the results for quantum clusters of significantly different sizes gives an estimate of the difference with the thermodynamic limit.
 The implementation of the hybrid  method can realistically involve only relatively small quantum clusters of 10-20 spins 1/2. Yet, precisely for this reason, the relative differences between these sizes are large. Therefore, if these differences do not lead to large deviations of the computed correlation functions, then the result should be viewed as reliable. For the lattices with not too small number of interacting neighbours, where purely classical calculations are expected to work well\cite{Elsayed-15},  the deviation between a purely classical calculation and a hybrid calculation with a small quantum cluster can already be sufficient for a reasonable estimate of the predictive accuracy.


\section{Tests}

\begin{figure*} \setlength{\unitlength}{0.1in}
\begin{picture}(70,50)(0,-13)
 
 \put(6, 18.5) {\includegraphics[width=2.8in]{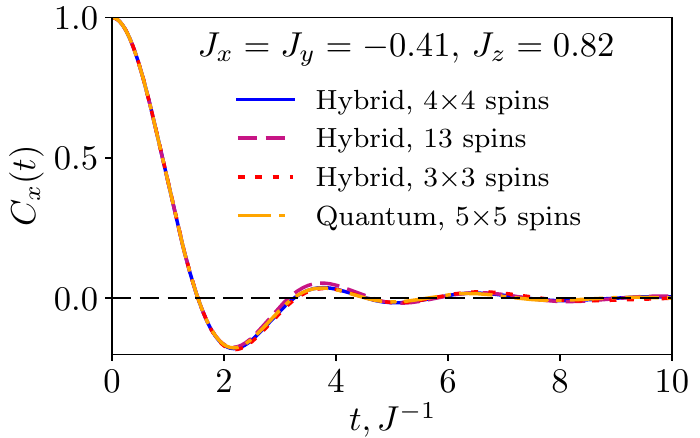}}
 \put(35, 18.5) {\includegraphics[width=2.8in]{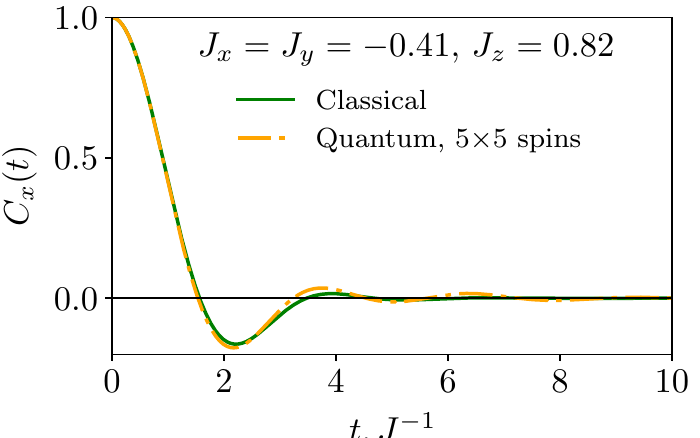}}
 
 \put(6, 0.5) {\includegraphics[width=2.8in]{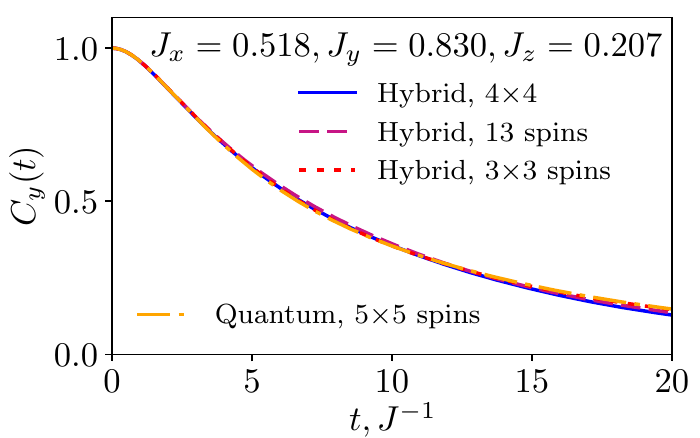}}
 \put(35,0.5) {\includegraphics[width=2.8in]{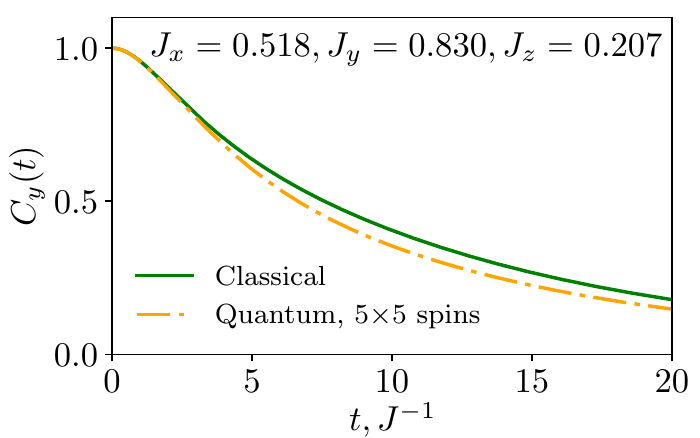}}
 
 \put(5.5, 36)  {\pmark a}
 \put(34.5, 36) {\pmark {a\tprime}}
 
 \put(5.5, 18)  {\pmark b}
 \put(34.5, 18) {\pmark {b\tprime}}
 
 \put(17,-13) {\includegraphics[width=3.5in]{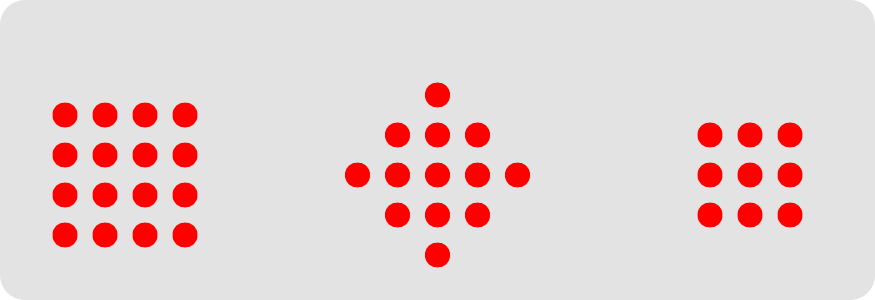}}
 
 \put(17,-3) {\parbox{1in}{\small4$\times$4 cluster}}
 \put(28.5,-3) {\parbox{1.2in}{\small 13-spin cluster}}
 \put(42, -3) {\parbox{1in}{\small3$\times$3 cluster}}
 
 \put(13.5,-1.5) {\pmark c}
\end{picture}
\caption{
Correlation functions $C_{\alpha} (t) $  for two-dimensional periodic lattices with nearest-neighbour interaction. The notations in (a,a',b,b') are the same as in Fig.~\ref{onedim}. For both hybrid and classical simulations, the full lattice size is $9\times9$. The shapes of quantum clusters for hybrid simulations are shown in (c).
        }
\label{twodim}
\end{figure*}
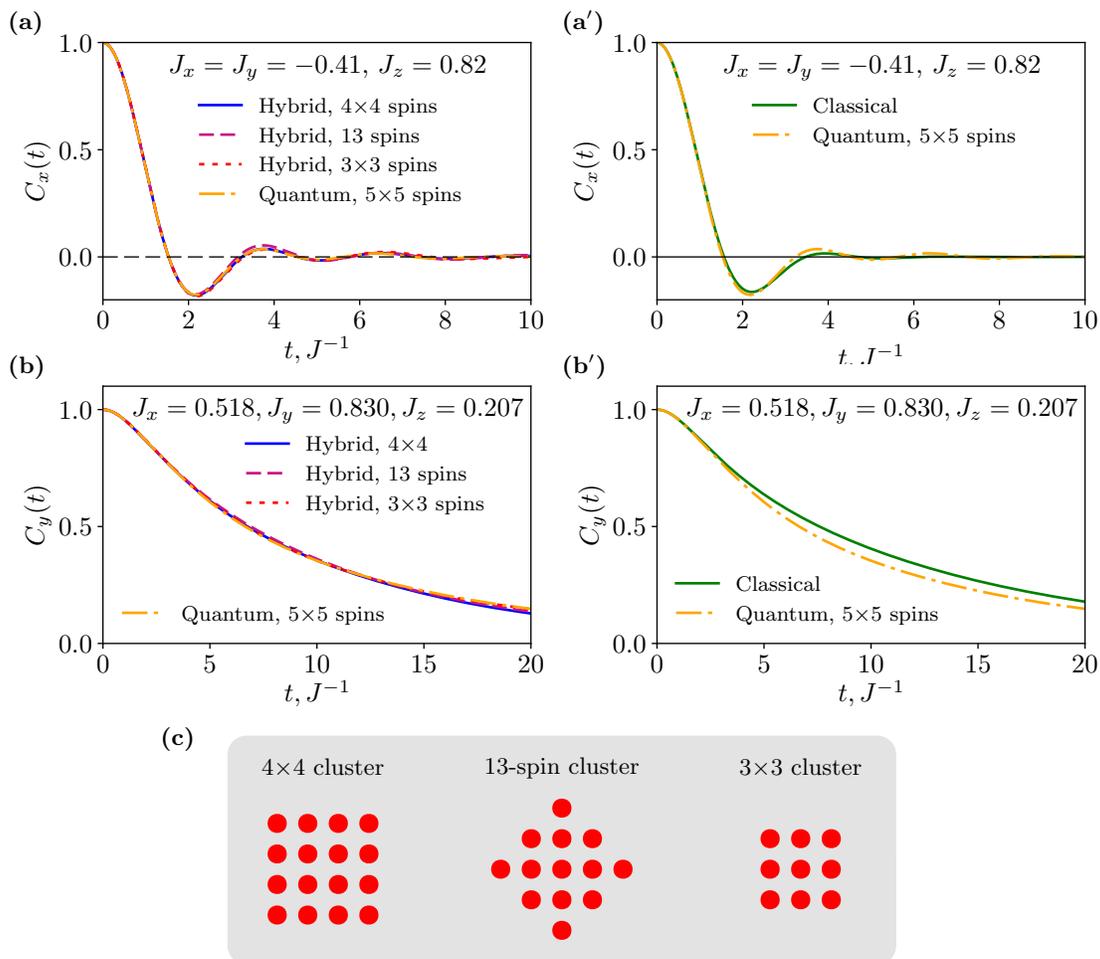

\begin{figure} \setlength{\unitlength}{0.1in}
\includegraphics[width=3.4in]{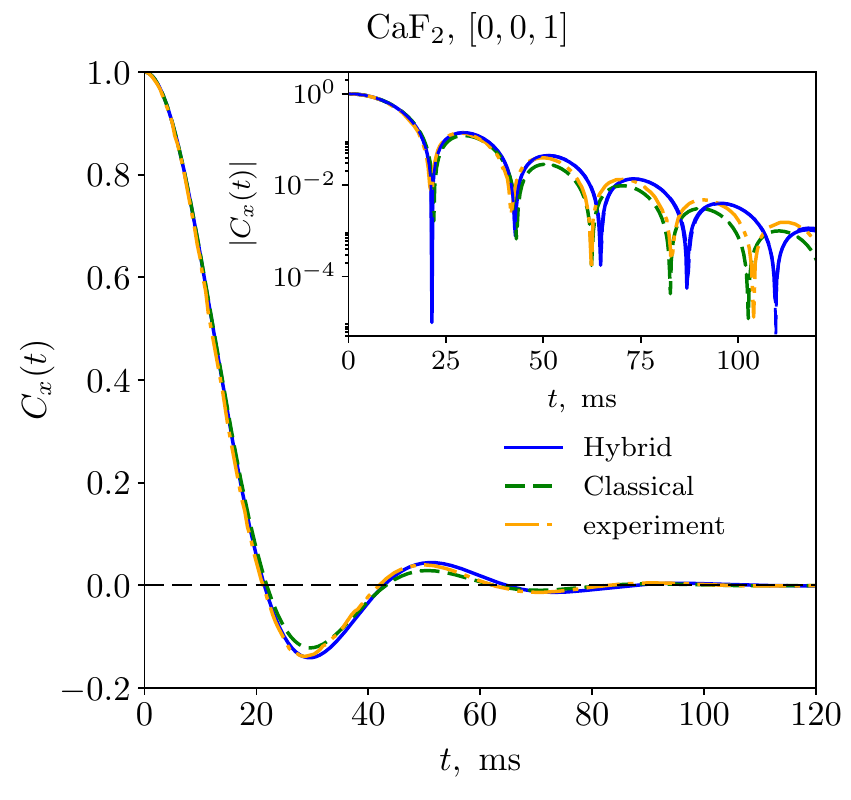}

\caption{FID in CaF$_2$ for external magnetic field $B_0$ along the $[001]$ crystal direction. Hybrid and classical simulations are compared with the experimental result of Ref.\cite{Engelsberg-74}. For both hybrid and classical simulations, the full lattice size is $9\times9\times9$. The quantum cluster in hybrid simulations was a chain extending along the $z$-axis ([001] crystal direction) through the entire lattice. The inset shows the semi-logarithmic plot of the same FID.}

\label{FIDfig}
\end{figure}

Our tests of the performance of the hybrid method for one-dimensional chains and two-dimensional square lattices of spins 1/2 are presented in Figs.~\ref{onedim} and \ref{twodim}, respectively. The lattices had nearest-neighbor interactions with coupling constants indicated in the figure legends. In the both figures, hybrid method's predictions are compared with the results of numerically exact quantum calculations for sufficiently large clusters. The cluster was considered ``sufficiently large'', when, in the time range of interest, the change of $C_{\alpha}(t)$ with the increase of the cluster size was negligible. The figures also include back-to-back comparison of hybrid simulations with purely classical simulations. More such tests can be found in the Supplemental Material.

For one-dimensional chains, the performance of the hybrid simulations in Figs.~\ref{onedim}(a,b)  is excellent. These two figures correspond to  typical situations when correlation functions $C_{\alpha}(t)$ decay not too slowly, i.e on the timescale of the order of $\tau_c$.  On the contrary, Fig.~\ref{onedim}(c) illustrates an atypical case, where the coupling constants and the axis $\alpha$ are chosen such that $C_{\alpha}(t)$ decays anomalously slowly. In this case, the hybrid method's prediction exhibits a clear discrepancy from the reference plot. Important, however, is the fact, also illustrated in Fig.~\ref{onedim}(c), that the internal estimate of the predictive accuracy based on the use of different quantum clusters within the hybrid method would anticipate the above discrepancy. We note here that the same accuracy estimate in Figs.~\ref{onedim}(a,b) is consistent with the observed excellent agreement with the reference plots. We further observe that, in all cases presented in Fig.~\ref{onedim}, the performance of the hybrid simulations is significantly better than that of the classical ones. 

Figure~ \ref{twodim} illustrates that, for two-dimensional lattices, hybrid simulations generally exhibit a very good performance, which is also noticeably better than that of the classical simulations, even though the latter is also reasonable --- consequence of the fact that the number of the interacting neighbors of each spin has increased in comparison with the one-dimensional case\cite{Elsayed-15}. 

For three-dimensional lattices, direct numerical calculation of reliable reference plots for sufficiently large quantum clusters is not feasible.
Therefore, we test the hybrid method by comparing its predictions with the NMR
FID  experiment \cite{Engelsberg-74} for $^{19}$F nuclei in the benchmark material CaF$_2$. 
These nuclei have spin 1/2, form a cubic lattice and interact {\it via} truncated magnetic-dipolar interaction (see Appendix~\ref{FID}). 
In Fig.~\ref{FIDfig}, we present the comparison between the experiment and the results of the hybrid and the classical simulations for magnetic field $B_0$ oriented along the $[0,0,1]$ crystal direction. In this case, classical simulations are known to lead to a good agreement with experiment --- consequence of the relatively large effective number of interacting neighbors. For the same reason, hybrid method was not expected to generate predictions very different from the classical ones irrespective of the choice of the quantum cluster within the method. Here we chose the quantum cluster in the form of a chain extending along the $z$-direction, because the nearest-neighbor coupling constant in that direction was the strongest, and hence we believed it was the best approach to preserve the resulting quantum correlations. As can be seen in Fig.~\ref{FIDfig}, the resulting performance of the hybrid simulations was somewhat better than that of the classical ones. More importantly, Fig.~\ref{FIDfig} illustrates the predictive uncertainty criterion formulated earlier, namely, that, for the lattices with large number of interacting neighbors, the deviation between the predictions of the two methods quantify the uncertainty of either of them. Indeed, hybrid and classical results diverge approximately at the same point where they start noticeably deviating from the experimental result. (See the Supplementary Information for similar tests with $B_0$ oriented along the $[0,1,1]$ and $[1,1,1]$ crystal directions.)
  

\section{Discussion} 

Overall, Figs.~\ref{onedim}, \ref{twodim} and \ref{FIDfig}, and the additional tests in the Supplementary Information illustrate that the hybrid method produces mostly very accurate predictions.  As we now explain, the rare situations where method's predictive accuracy  is limited can be understood from the analysis of the asymptotic long-time behavior of $C_\alpha(t)$.


There exists substantial experimental\cite{Engelsberg-74,Morgan-08,Sorte-11,Meier-12} and numerical\cite{Fabricius-97,Fine-03,Elsayed-13,Elsayed-15} evidence, also supported by theoretical arguments\cite{Borckmans-68,Fine-04,Fine-05}, that, despite widely varying shapes of correlation functions $C_\alpha(t)$,  their long-time behaviour in non-integrable systems has universal form
\begin{equation}
 C_\alpha(t)\cong  e^{-\gamma t} \ \  \text{or} \ \ C_\alpha(t)\cong  e^{-\gamma t}\cos{(\omega t+ \phi)},
 \label{relax_modes}
\end{equation}
where $\gamma$ and $\omega$ are constants of the order of $1/\tau_c$. 
The asymptotic behavior(\ref{relax_modes}) represents the slowest-decaying relaxational mode of the system\cite{Fine-04}. Typically, it  becomes dominant after time of the order of several $\tau_c$.  Therefore, if one manages to accurately compute $C_\alpha(t)$ over the above initial time interval, then a good overall accuracy is assured. This is what the hybrid method achieves in a typical setting.

On the basis of the above consideration, one can anticipate that the hybrid method would predict the asymptotic time constants $\gamma$ and $\omega$ with {\it absolute} uncertainty $\epsilon/\tau_c$, where $\epsilon$ is a number significantly smaller than 1. Yet, such an uncertainty may lead to noticeable 
discrepancies in two problematic cases\cite{Fine-04}: In the first of them, the slowest relaxational mode is characterized by $\gamma\ll1/\tau_c$, and hence the {\it relative} uncertainty of predicting $\gamma$ may be large [cf. Fig.~\ref{onedim}(c)]. In the second problematic case, the asymptotic behavior is characterized by an accidental competition between two slowest relaxation modes with exponential decay constants $\gamma_1$ and $\gamma_2$ such that $|\gamma_2-\gamma_1| \ll1/\tau_c$. As a result, the long-time behavior can be significantly distorted in an approximate calculation. 
The above analysis further implies that the competition between two relaxational modes in the long-time regime is accompanied by the increased sensitivity of direct quantum simulations to the  size and shape of the quantum cluster, which, in turn, prevented us from conclusively testing the hybrid method in the presence of two-mode competition [see Supplementary Information]. 

We, finally, remark that there exists a straightforward extension of the present method, where, instead of dividing the simulated lattice into a quantum cluster and a classical environment, one can divide it into computationally manageable quantum clusters coupled to each other via local fields of form (\ref{backaction}) obtained from the quantum mechanical expectations values of spin operators within each cluster. Our preliminary investigations have not revealed any clear computational advantages of the latter approach in comparison with the hybrid method.


\section{Conclusions}

In conclusion, we proposed a hybrid quantum-classical method of simulating high-temperature dynamics of nuclear spins in solids. The method exhibits excellent overall performance for quantum spin lattices of different dimensions and with different interactions. It comes with a long-sought internal estimate of the predictive accuracy, which was validated in each of the large number of tests we have performed. The method can, therefore, be used to make reliable predictions of NMR spin-spin relaxation in various materials with the goal of extracting unknown microscopic information, such as the distances between nuclei or the mechanisms of coupling between them.

\acknowledgments

The authors are grateful to O. Lychkovskiy and A. Rozhkov for discussions. This work was supported by a grant of the Russian Science Foundation (Project No. 17-12-01587).



{\appendix

\section{Classical simulations}
\label{classical}

The equations of motion for  classical spins are obtained from Hamiltonian (\ref{cl_ham}) with the help of Poisson brackets\cite{deWijn-13} $\{s_m^\alpha,s_n^\beta\}_P = \delta_{mn} \, e_{\alpha\beta\gamma} \, s_m^\gamma$, which gives:
\begin{equation}
 \dot{\bf s}_m =  \{{\bf s}_m, {\BH}_\BC \}_P = {\bf s}_m\times ({\bf h}_m^{\BC\BC} + {\bf h}_m^{\BQ\BC}),
 \label{cl_eq}
\end{equation}
where
\begin{equation}
    {\bf h}^{\BC\BC}_m = -\lsum_{n\in\mathcal{C}}\left(\begin{array}{c}
                                             J_{m,n}^x s_n^x \\
                                             J_{m,n}^y s_n^y \\
                                             J_{m,n}^z s_n^z \\
                                            \end{array}
\right).\label{clfield}\\
\end{equation}

The infinite temperature state is characterized by completely random orientations of classical spins. Therefore, the initial spin vectors $\{{\bf s}_m(0)\}$ were generated as radius-vectors of points randomly sampled on a sphere of radius $\sqrt{S(S+1)} = \sqrt{3}/2$ with uniform probability distribution.
The length of classical spin vectors $\sqrt{S(S+1)}$ guarantees that the characteristic time $\tau_c$ is the same for classical and quantum lattices. It also guarantees the equality of the second moments $M_2 \equiv -C^{\prime\prime}_\alpha(0)/C_\alpha(0)$ for the two lattices. With such a choice, correlation functions corresponding to purely quantum and purely classical lattices are known to become very close to each other\cite{Jensen-73,Tang-92,Elsayed-15}, when the effective number of interacting neighbors of each spin
\begin{equation}
n_{\text{eff}}  \equiv 
\frac{
\left[  
\sum_n \left( {J^x_{mn}}^{\!\!\!\! 2}  +   {J^y_{mn}}^{\!\!\!\! 2}   +   {J^z_{mn}}^{\!\!\!\! 2}  \right) 
 \right]^2
 }
 {  
\sum_n  \left( {J^x_{mn}}^{\!\!\!\! 2}  +   {J^y_{mn}}^{\!\!\!\! 2}   +   {J^z_{mn}}^{\!\!\!\! 2}    \right)^2
 }
 \label{neff}
 \end{equation}
 is greater than four\cite{Elsayed-15}.  It was also shown analytically in Ref.\cite{Lundin-77}, that, in the limit of infinite number of interacting neighbours, the two kinds of correlation functions are supposed to become identical.


\section{Quantum Simulations}
\label{quantum}

The dynamics of quantum clusters was simulated by the method of direct time integration of the Schr\"odinger equation
\begin{equation}
 \cfrac{d}{dt}|\psi(t)\rangle = -i\BH|\psi(t)\rangle.
 \label{schroedinger}
\end{equation}
without the complete diagonalization of the Hamiltonian\cite{Elsayed-13}. In comparison with the latter, the direct integration allows one to treat larger quantum clusters numerically exactly, because it does not require one to store in the computer memory either density matrices or unitary transformations, which are dense $N\times N$ matrices. Instead, only the wave function vector and the sparse Hamiltonian matrix are stored.

Each simulation started from a randomly sampled pure quantum state (a superposition of eigenstates) representing the infinite temperature.  These initial states were generated as 
  \begin{equation}
   |\psi_\text{rand}(0)\rangle  = \lsum_{k=1}^N a_k e^{i\varphi_k}|k\rangle,
   \label{random}
  \end{equation}
where ${|k\rangle}$ was a full orthonormal basis,  $a_k$ real numbers distributed according to the probability distribution
  \begin{equation}
   P(a_k^2) = N\exp{(-N a_k^2)},
  \end{equation}
  and $\varphi_k$ random phases selected from interval $[0,2\pi)$ \cite{Elsayed-13}. The wave functions $|\psi_\text{rand}(0)\rangle$ were then
normalized.  

Once $|\psi(t)\rangle$ is obtained, one can compute the quantum expectation value $\langle\psi(t)|\sum_m{S}_m^\alpha|\psi(t)\rangle$ and then use it to obtain the correlation function $C_{\alpha}(t)$ (see below).



\section{Numerical integration of quantum and classical equations of motion}
\label{numerical}

In hybrid simulations, the dynamical  equations (\ref{cl_eq}, \ref{schroedinger}) are integrated jointly using explicit Runge-Kutta scheme of 4-th order with fixed time step of $2^{-7}\,J^{-1}$, or, in some cases, $2^{-6}\,J^{-1}$ (to speed up the calculations). The time unit $J^{-1}$ is defined as follows: for one-dimensional and two-dimensional lattices, $J = \sqrt{J_x^2 + J_y^2 + J_z^2} $, where $J_x$, $J_y$, $J_z$ are the nearest-neighbor coupling constants; for the three-dimensional CaF$_2$ lattice, $J = g^2 \hbar^2/a_0^3$, where $g$ is the gyromagnetic ratio and $a_0$ is the cubic lattice period, both appearing in Eq.(\ref{dipolar}) below.  The choice of the time step is discussed in Refs.\cite{Elsayed-13,Elsayed-15}. Purely classical or purely quantum simulations are performed as the appropriate limit of the hybrid simulations. The numbers of computational runs (realizations of the time evolution of the system starting from randomly chosen initial conditions) from which the plotted correlation functions were extracted is given in the Supplementary Material.


\section{Representations of correlation functions}
\label{representations}

For purely classical systems, we extracted equilibrium correlation functions $C_\alpha(t)$ from the equilibrium noise of the quantity of interest $M_\alpha(t) = \sum_m s_m^{\alpha}(t)$ using the following definition:
  \begin{equation}
   C_\alpha(t) = \BN\cdot\left[\cfrac{1}{T_{max}}\int\limits_{0}^{T_{max}} d\tau {M_\alpha(\tau+t)M_\alpha(\tau)}\right]_{i.c.},
   \label{cl_cor}
  \end{equation}
  where $\BN$ is normalization constant and $[...]_{i.c.}$ denotes averaging over the infinite-temperature ensemble of initial conditions. The time $T_{max}$ was chosen to be sufficiently large ($T_{max}\gg\tau_c$, $T_{max}\gg t$). In principle, if the system is ergodic and the limit $T_{max} \to \infty$ is taken, then the averaging over the initial conditions is not necessary. In practice, however, given the unclear ergodization timescales, we perform the additional averaging over initial conditions both as a consistency check, and as a way to improve the efficiency of the averaging procedure.
  
For purely quantum systems, the correlation function of interest is, at first sight, defined differently, namely, as  
$C_\alpha(t) \simeq \text{Tr} \left\{ e^{i {\BH}_\BQ t} M_\alpha e^{-i {\BH}_\BQ t} M_\alpha   \right\} $, where $M_\alpha = \sum_m S_m^{\alpha}$ is a quantum-mechanical operator. It was, however, proven in Ref.\cite{Elsayed-13}, that one can obtain the result of the above quantum trace calculation with the help of formula (\ref{cl_cor}), where classical projections $M_\alpha(t)$ are replaced by  quantum-mechanical expectation values $M_\alpha(t) = \langle\psi(t)|\lsum_{m\in\BQ} {S}_m^\alpha|\psi(t)\rangle$ associated with the time evolution of a randomly chosen wave function (\ref{random}). The amplitude of the resulting quantum noise of $M_\alpha(t)$ is, however, smaller than that of the classical counterpart by factor $1/\sqrt{N}$.  

At the level of the basic idea, our method of hybrid simulations compensates the above amplitude mismatch by redefining $M_\alpha(t)$ with the help of Eq.(\ref{total_spin}), and then obtaining $C_\alpha(t)$ using Eq.(\ref{cl_cor}) with the newly defined $M_\alpha(t)$. 
However, in the final application of the method, we introduce an additional technical modification aimed at reducing the effect of the quantum-classical border. Namely, we use the fact that, due to the translational invariance of the  original quantum problem, the correlation function of interest can be reexpressed as 
$C_\alpha(t) \simeq \text{Tr} \left\{ e^{i {\BH}_\BQ t} M_\alpha e^{-i {\BH}_\BQ t} S_m^\alpha   \right\} $, where $S_m^\alpha$ is the $\alpha$th projection operator of any spin on the lattice. Moreover, $S_m^\alpha$ in this expression can be further replaced by the sum
$M_\alpha' = \lsum_{m\in\BQ'} {S}_m^\alpha$ over any subset $\BQ'$ of spins on the lattice, which, therefore, we can choose at our discretion.
The presence of the quantum-classical border in the hybrid simulations breaks the translational invariance of the system, thereby making different choice of $M_\alpha'$ nonequivalent from the viewpoint of the approximation error.  We minimize this error, by choosing subset $\BQ'$  to consist of one or several equivalent quantum spins which are furthermost from the quantum-classical border. 

Finally, we combine all the above relations to arrive at the expression for the correlation function actually used in our hybrid simulations:
 \begin{equation}
   C_\alpha(t) = \BN\cdot\left[\cfrac{1}{T_{max}}\int\limits_{0}^{T_{max}} d\tau {M_\alpha(\tau+t)M_\alpha'(\tau)}\right]_{i.c.},
   \label{cl_cor_prime}
  \end{equation}
where $M_\alpha(t)$ is given by Eq.(\ref{total_spin}), and $M_\alpha' = \sqrt{N}\cdot\langle\psi(t)|\lsum_{m\in\BQ'} {S}_m^\alpha|\psi(t)\rangle$. For each set of initial conditions, we integrated the dynamical equations (\ref{cl_eq}, \ref{schroedinger}) up to time 
$T_{max} \sim 10 T_0$, where $T_0$ is the maximum time $t$ in Eq.(\ref{cl_cor_prime}) for which the correlation function $C_\alpha(t)$ was to be computed. The number of initial conditions was then chosen sufficiently large to make the resulting statistical uncertainty of $C_\alpha(t)$ negligible on the scale of the resulting plots.

We tested hybrid simulations for one- and two-dimensional lattices by comparing hybrid results with purely quantum simulations of larger spin-1/2 clusters, for which the direct integration of the Schr\"odinger equation could be implemented numerically   (typically, up to 25 spins 1/2).   For this, we used yet another representation of the  correlation function\cite{Zhang-07,Elsayed-13}:
\begin{equation}
 C_\alpha(t) = \BN\cdot \left[\langle\psi(t)|\sum_m{S}_m|\psi_{aux}(t)\rangle\right]_{i.c.},
\end{equation}
where $ |\psi(t)\rangle $ is obtained {\it via} direct integration starting from a randomly selected $|\psi(0)\rangle$, while the 
$|\psi_{aux}(t)\rangle$ is obtained via the direct integration of the unnormalized auxiliary initial wave function $\sum_m{S}_m |\psi(0)\rangle$.
Such a method is more efficient than the one involving formula (\ref{cl_cor_prime}), because, for larger clusters, it requires the direct integration of only two wave functions over time $T_0$ (much less than $T_{max}$) to obtain $C_\alpha(t)$ with accuracy $1/\sqrt{N}$.
So far, however, we were not able to incorporate this method into a hybrid simulation scheme.


\section{Free induction decay in CaF$_2$}
\label{FID}

 The FID experiments in solids measure the relaxation of the total nuclear magnetization transverse to a strong magnetic field ${\bf B}_0$. The relaxation is caused by the magnetic dipolar interaction between nuclear spins averaged over the fast Larmor precession induced by ${\bf B}_0$. The effective interaction Hamiltonian in the Larmor rotating reference frame has form (\ref{Ham}) with coupling constants:
\begin{equation}
 J^z_{i,j} = -2J^x_{i,j} = -2J^y_{i,j} = \cfrac{g^2\hbar^2(1-3 \cos^2{\theta_{ij}})}{|{\bf r}_{ij}|^3},\label{dipolar}
\end{equation}
where the $z$-axis is chosen along the direction of ${\bf B}_0$,  ${\bf r}_{ij}$ is the vector connecting lattice sites $i$ and $j$, $\theta_{ij}$ is the angle between ${\bf r}_{ij}$ and ${\bf B}_0$, $g$ is the gyromagnetic ratio of nuclei. The measured FID signal is proportional to $C_x(t)$ given by Eq.(\ref{corf1}).

In CaF$_2$, $^{19}$F nuclei form a cubic lattice with period $a_0=2.72$~\AA. Their gyromagnetic ratio is $g = 2.51662\cdot 10^8 \text{ rad s}^{-1}\text{ T}^{-1}$.

}

\clearpage
\pagebreak

{\centerline{\bf\large SUPPLEMENTARY INFORMATION}}

\setcounter{figure}{0}

\renewcommand{\thefigure}{S\arabic{figure}}

\setcounter{equation}{0}

\renewcommand{\theequation}{S\arabic{equation}}

\setcounter{table}{0}

\renewcommand{\thetable}{S\arabic{table}}

\setcounter{section}{0}
\renewcommand{\thesection}{S\Roman{section}}

\section{Suppression of the expectation values of quantum operators\\ by factor $1/\sqrt{N}$}

Let us consider a cluster of $N_{\BQ}$ spins 1/2 with the dimension of the Hilbert space $N= 2^{N_{\BQ}}$. Let us further consider quantum operator $A$, which has infinite-temperature average $ \langle A \rangle \equiv {1 \over N} \text{Tr} A = 0$ and the variance $ \langle A^2 \rangle \equiv {1 \over N} \text{Tr} A^2 \equiv A_{\text{rms}}^2 $. This can be the operator of local field, or the projection of an individual spin, or the operator of the total spin polarization. Here we show that, for a wave function $|\psi \rangle$ randomly sampled in the Hilbert space of the cluster according to prescription (\ref{random}),
 \begin{equation}
 \langle \psi | A  | \psi \rangle \sim A_{\text{rms}}/\sqrt{N},
 \label{Aav}
 \end{equation}
 The intuitive explanation of this fact is based on the notion of quantum parallelism\cite{Pastawski-08}. Namely, the expectation value $\langle \psi | A  | \psi \rangle$ can be thought of as the average over $N$ independent realizations of the state of the system, where factor $1/\sqrt{N}$ reflects the quality of statistical averaging. 
 
 Formally, the relation (\ref{Aav}) can be proven\cite{Elsayed-13} by demonstrating that  
  \begin{equation}
 \sqrt{\left[ \langle \psi | A  | \psi \rangle^2 \right]_{\psi}} = A_{\text{rms}}/\sqrt{N},
 \label{Aav1}
 \end{equation}
 where $[...]_{\psi}$ denotes the average over all possible realizations of $\psi$.
Using representation (\ref{random}), one can express such an average in terms of integrals over expansion coefficients $c_k \equiv a_k e^{i \varphi_k}$:
\begin{multline}
 \left[ {\cal F}\left(  \left\{ a_k, \varphi_k \right\} \right) \right]_{\psi} \equiv \\ \* \left[ \Pi_{k}\lint_{0}^{+\infty} d(a_k)^2\int\limits_{0}^{2\pi} \cfrac{d\varphi_k}{2\pi} N exp(-Na_k^2)\right] {\cal F}\left(  \left\{ a_k, \varphi_k \right\} \right)
\end{multline}

With this definition, one can obtain\cite{Elsayed-13}:
\begin{align}
 [c_k^*c_m]_{\psi} & = \delta_{k,m}/N \\
 [c_k^*c_m c_n^* c_l]_{\psi} & = (\delta_{k,m}\delta_{n,l}+\delta_{k,l}\delta_{m,n})/N^2
\end{align}
As a result, 
\begin{equation}
 [\langle\psi|A|\psi\rangle]_{\psi} = \lsum_{m,n} [c^*_mc_n]_{\psi} A_{mn} =  \tr{[A]} /N= 0,
 \label{rel1}
 \end{equation}
 where $A_{mn}$ are the matrix elements of $A$, and
 \begin{multline}
 [\langle\psi|A|\psi\rangle^2]_{{\psi}} = \lsum_{k,l,m,n} [c^*_kc_lc^*_mc_n]_{\psi} A_{kl}A_{mn} = \\ \* =\cfrac{\tr{[A^2]}}{N^2} + \cfrac{\tr^2{[A]}}{N^2} = \cfrac{A_{\text{rms}}^2}{N} ,
  \label{rel2}
\end{multline}
which gives Eq.(\ref{Aav1}).

Now we apply the above general result to the operator of local magnetic field of a quantum spin lattice
  \begin{equation}
      {\bf h}_i = -\lsum_{j\neq i}\left(\begin{array}{c}
                                             J_{ij}^x S_j^x \\
                                             J_{ij}^y S_j^y \\
                                             J_{ij}^z S_j^z \\
                                            \end{array}
\right).
\end{equation}
The root-mean-squared value of ${\bf h}_i$ is defined as
\begin{equation}
 h_{rms} \equiv 
 \sqrt{ 
 {1 \over N}
 \sum_{j\neq i,\alpha} \left( {J_{ij}^\alpha}^2 \tr{[{S_j^\alpha}^2]}\right)
 }.
\end{equation}
(The characteristic time of lattice dynamics $\tau_c$ given in the main text is obtained as $1/h_{rms}$.)

If we consider quantum expectation value $\langle\psi|{\bf h}_i |\psi\rangle$ for a random quantum state, then its root-mean-squared value is
\begin{multline}
 \parallel\langle\psi|{\bf h}_i |\psi\rangle\parallel_{rms} = \sqrt{\left[\langle\psi|{\bf h}_i|\psi\rangle^2\right]_{\psi}} = \\ \* =
 \sqrt{\lsum_{j\neq i,l\neq i,\alpha} J^\alpha_{ij}J^\alpha_{il} \left[S_j^\alpha S_l^\alpha \right]_{\psi}}.
\end{multline}
Using Eqs.(\ref{rel1}) and the fact that $\left[S_j^\alpha S_l^\alpha \right]_{\psi}=0$ for $j\neq l$, we obtain:
\begin{equation}
 \parallel\langle\psi|{\bf h}_i|\psi\rangle\parallel_{rms} = \sqrt{\lsum_{j\neq i,\alpha}{J_{ij}^\alpha}^2 \cfrac{\tr{[{S_j^\alpha}^2]}}{N^2}} = \cfrac{h_{rms}}{\sqrt{N}}.
\end{equation}

\section{Finite-size analysis for quantum clusters used in Figs. \ref{onedim} and \ref{twodim}}

\begin{figure*} \setlength{\unitlength}{0.1in}
\begin{picture}(70,39)
 
 \put(6, 20.5) {\includegraphics[width=2.8in]{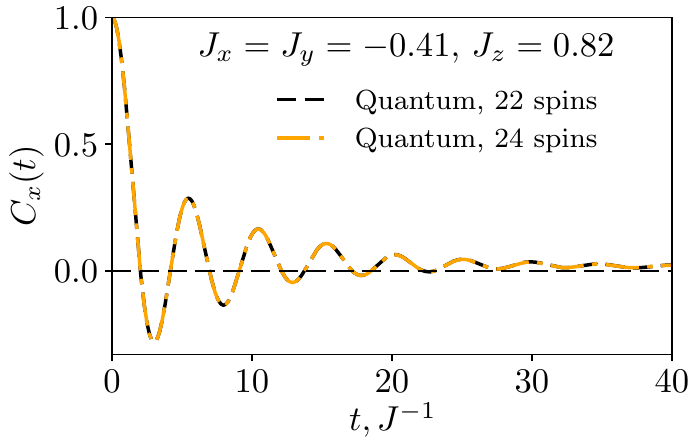}}
 \put(35, 20.5) {\includegraphics[width=2.8in]{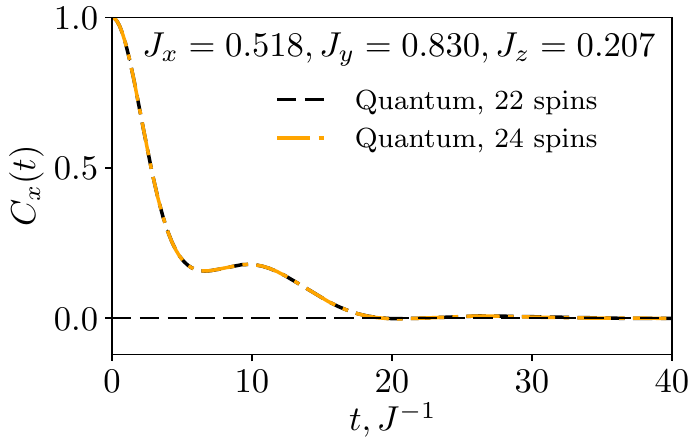}}
 
 \put(20, 0.5) {\includegraphics[width=2.8in]{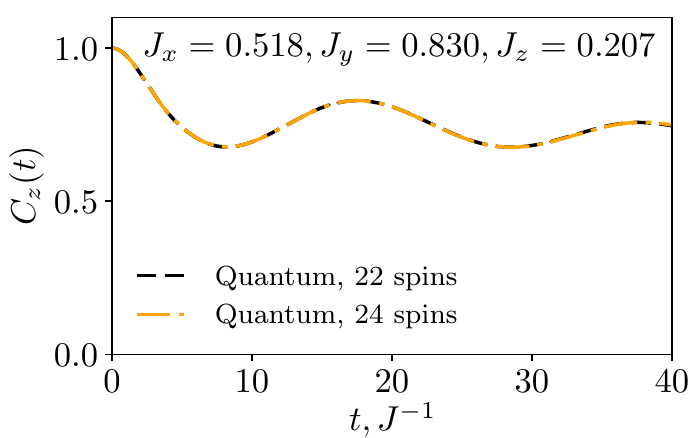}}
 
 \put(5.5, 38)  {\pmark a}
 \put(34.5, 38) {\pmark b}
 
 \put(19.5, 18)  {\pmark c}
\end{picture}
\caption{
Size dependence of correlation functions $C_{\alpha} (t) $  for one-dimensional periodic chains with nearest-neighbour interactions obtained from direct quantum calculations. The interaction constants are the same as in Fig.~\ref{onedim}. The present figure illustrates that quantum reference plots used in Fig.~\ref{onedim} represent the thermodynamic limit.
        }
\label{onedim_ref}
\end{figure*}

\begin{figure*} \setlength{\unitlength}{0.1in}
\begin{picture}(70,19)
 
 \put(6, 0.5) {\includegraphics[width=2.8in]{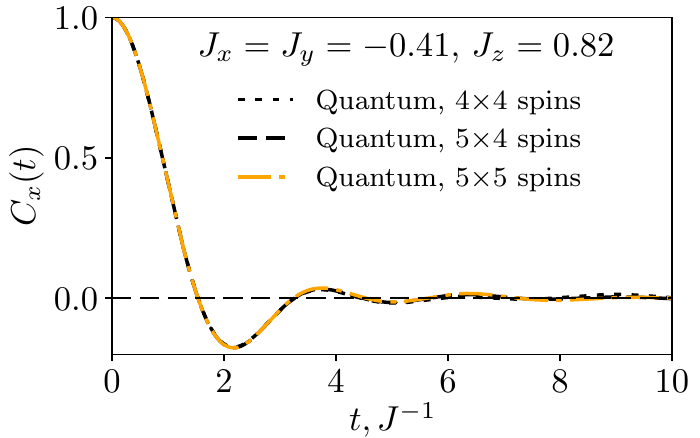}}
 \put(35, 0.5) {\includegraphics[width=2.8in]{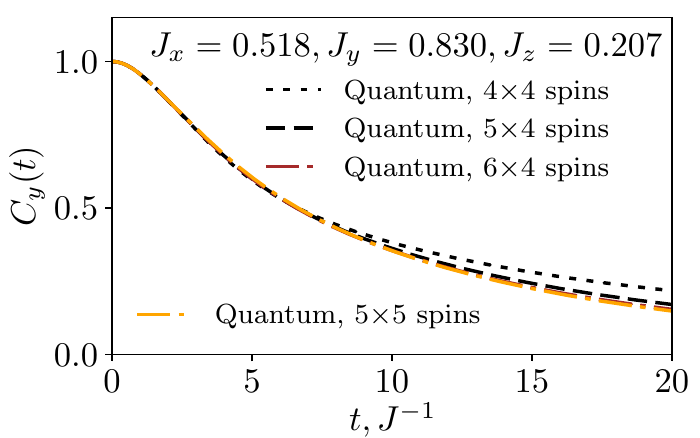}}
 
 \put(5.5, 18)  {\pmark a}
 \put(34.5, 18) {\pmark b}
\end{picture}
\caption{
Size dependence of correlation functions $C_{\alpha} (t) $  for two-dimensional periodic chains with nearest-neighbour interactions obtained from purely quantum simulations. The interaction constants are the same as in Fig.~\ref{twodim}. These plots illustrates that quantum results used in Fig.~\ref{twodim} as references represent the thermodynamic limit.
        }
\label{twodim_ref}
\end{figure*}

In Figs. \ref{onedim} and \ref{twodim} of the main article, we tested the performance of the hybrid method by comparing its predictions with the results obtained by direct calculations for purely quantum clusters of finite sizes. Here, in Figs.~\ref{onedim_ref} and \ref{twodim_ref}, we present the dependence of those results on the size of quantum clusters for one- and two-dimensional lattices respectively. These tests reveal that the correlation functions obtained for several cluster sizes coincide with a good accuracy, which, in turn, indicates that the respective plots represent the correlation functions of interest in the thermodynamic (infinite-cluster) limit.

\section{Additional tests of the hybrid method}

\begin{figure*} \setlength{\unitlength}{0.1in}
\begin{picture}(70,37)
 
 \put(6, 18.5) {\includegraphics[width=2.8in]{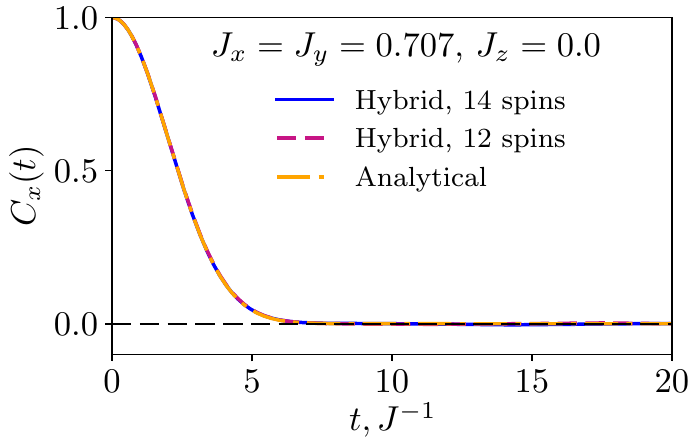}}
 \put(35, 18.5) {\includegraphics[width=2.8in]{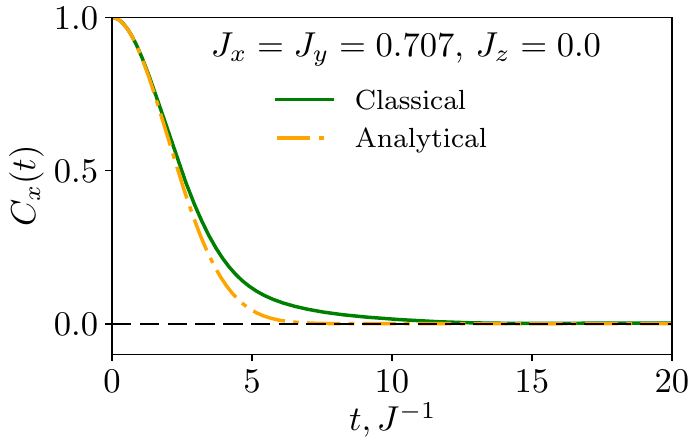}}
 
 \put(6, 0.5) {\includegraphics[width=2.8in]{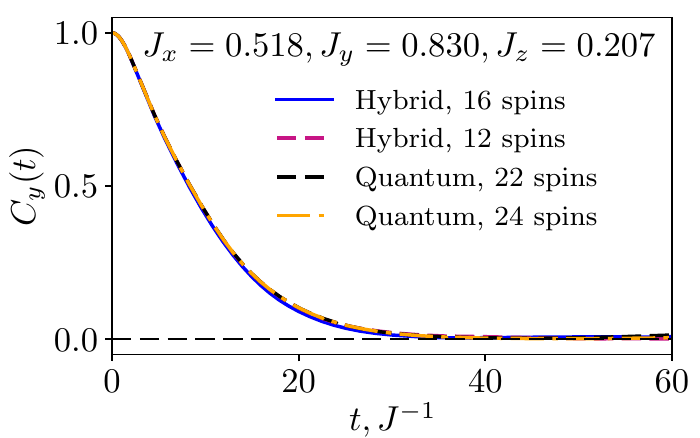}}
 \put(35,0.5) {\includegraphics[width=2.8in]{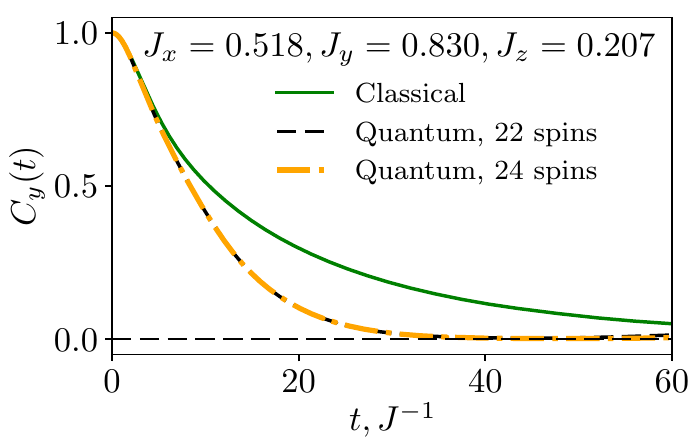}}
 
 \put(5.5, 36)  {\pmark a}
 \put(34.5, 36) {\pmark {a\tprime}}
 
 \put(5.5, 18)  {\pmark b}
 \put(34.5, 18) {\pmark {b\tprime}}
\end{picture}
\caption{Correlation functions $C_{\alpha} (t) $  for one-dimensional periodic chains with nearest-neighbour interactions (additional plots). The notations in are the same as in Fig.~\ref{onedim} of the main text.
For both hybrid and classical simulations, the full lattice size is $92$. Lines in (a,a') labeled as ``Analytical''  are Gaussians that represent the  analytical result  for the spin-1/2 $XX$ chain in the thermodynamic limit [U. Brandt and K. Jacoby, Z. Phys. B {\bf 25}, 181 (1976)].         }
\label{onedim_extra}
\end{figure*}

\begin{figure*} \setlength{\unitlength}{0.1in}
\begin{picture}(70,43)(0,-6)
 
 
 \put(6, 18.5) {\includegraphics[width=2.8in]{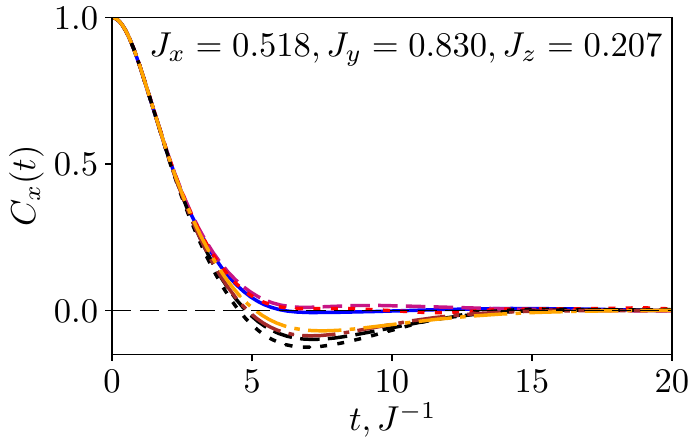}}
 \put(35,18.5) {\includegraphics[width=2.8in]{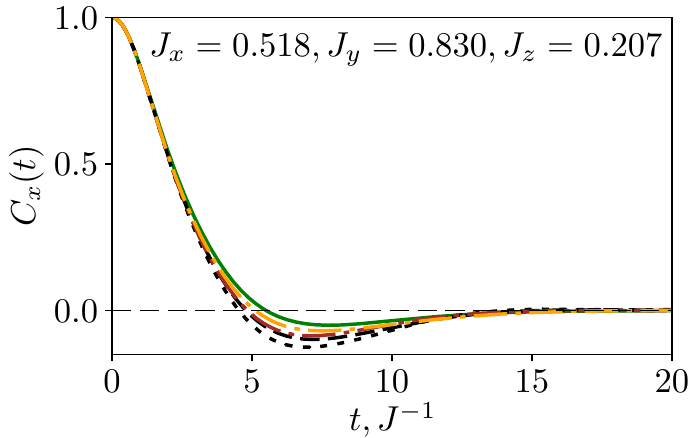}}
 
 \put(6, 0.5) {\includegraphics[width=2.8in]{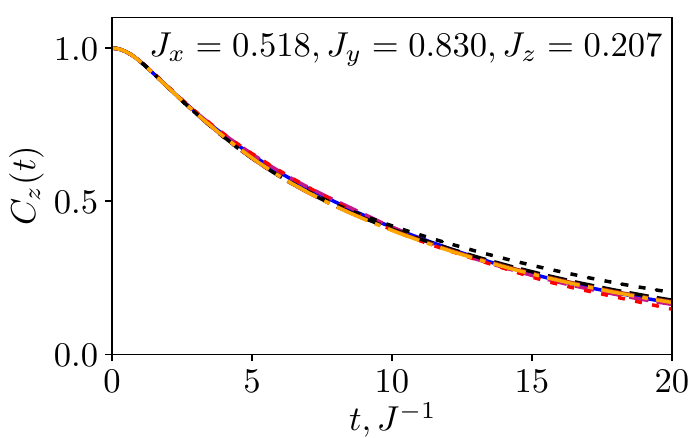}}
 \put(35,0.5) {\includegraphics[width=2.8in]{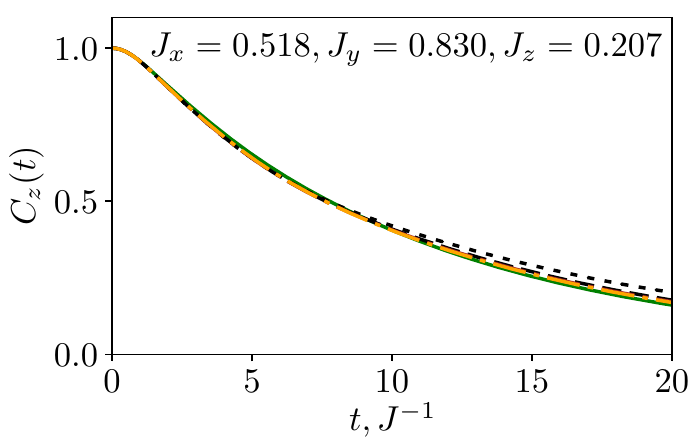}}
 
 \put(8,-6)    {\includegraphics[width=5.5in]{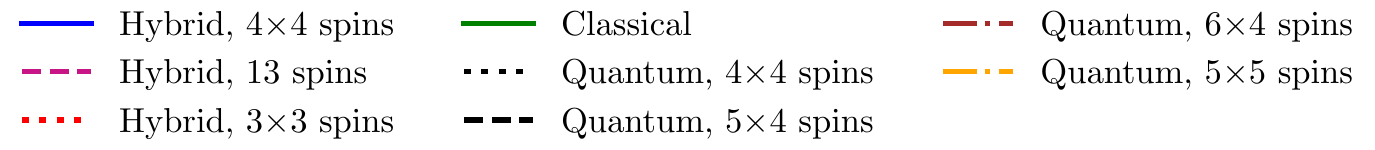}}

 \put(5.5, 36)  {\pmark a}
 \put(34.5, 36) {\pmark {a\tprime}}

 \put(5.5, 18)  {\pmark b}
 \put(34.5, 18) {\pmark {b\tprime}}
\end{picture}
\caption{Correlation functions $C_{\alpha} (t) $  for two-dimensional periodic lattices with nearest-neighbour interactions (first set of additional plots). The notations are the same as in Fig.~\ref{onedim} of the main text. For both hybrid and classical simulations, the full lattice size is $9\times9$. The shapes of quantum clusters for hybrid simulations are shown in Fig.~\ref{onedim}(c).
        }
\label{twodim_extra_1}
\end{figure*}

\begin{figure*} \setlength{\unitlength}{0.1in}
\begin{picture}(70,79)(0,-6)
 \put(6, 54.5) {\includegraphics[width=2.8in]{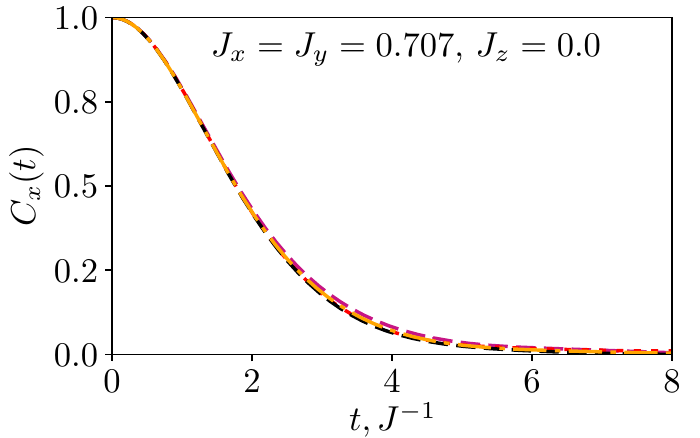}}
 \put(35, 54.5) {\includegraphics[width=2.8in]{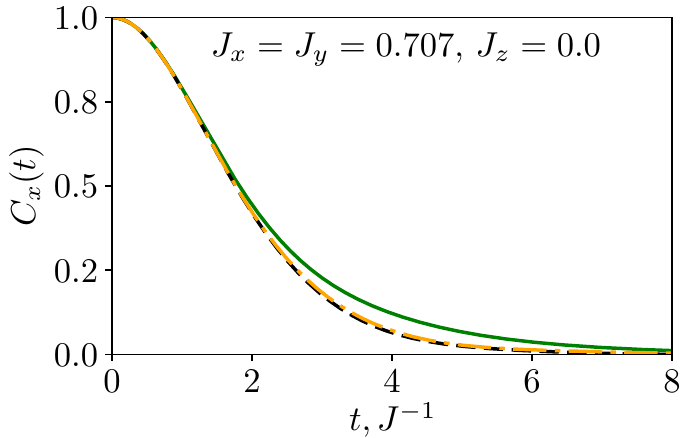}}
 
 \put(6, 36.5) {\includegraphics[width=2.8in]{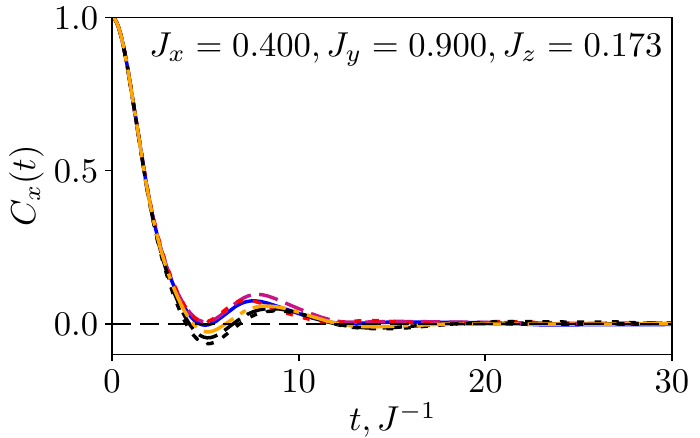}}
 \put(35,36.5) {\includegraphics[width=2.8in]{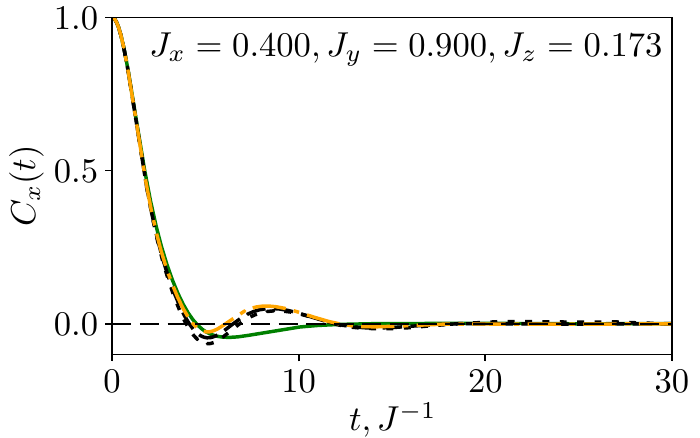}}
 
 \put(6, 18.5) {\includegraphics[width=2.8in]{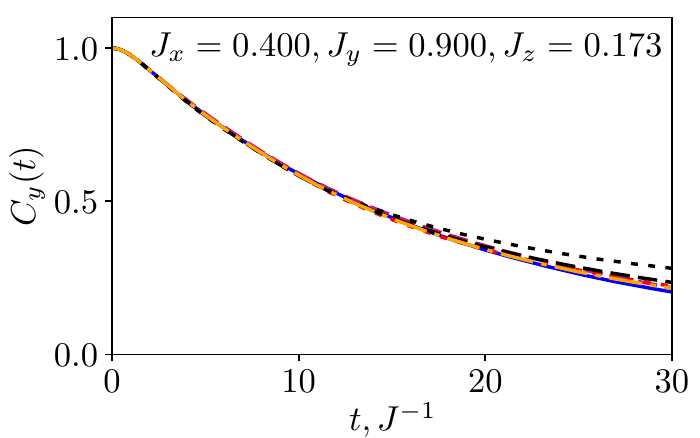}}
 \put(35,18.5) {\includegraphics[width=2.8in]{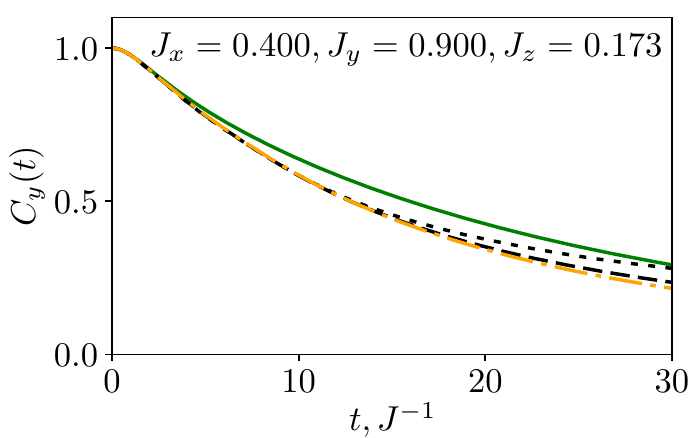}}
 
 \put(6, 0.5) {\includegraphics[width=2.8in]{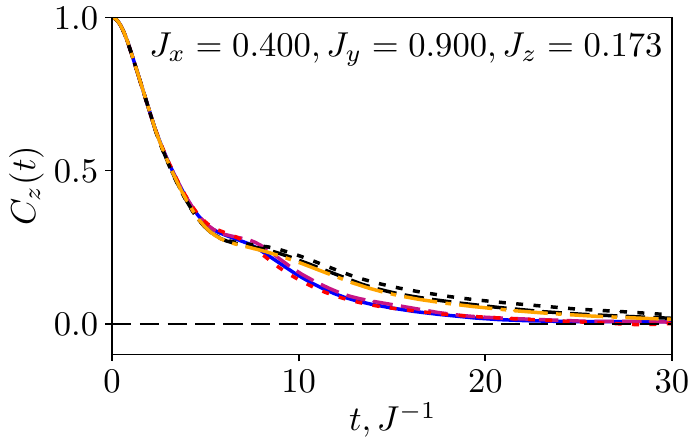}}
 \put(35,0.5) {\includegraphics[width=2.8in]{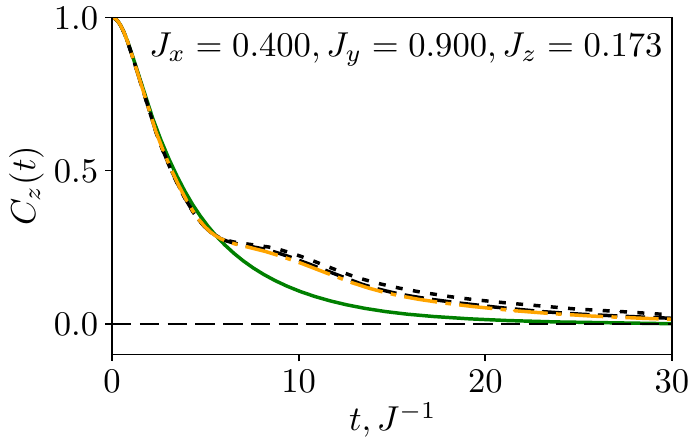}}
 
  \put(8,-6)    {\includegraphics[width=5.5in]{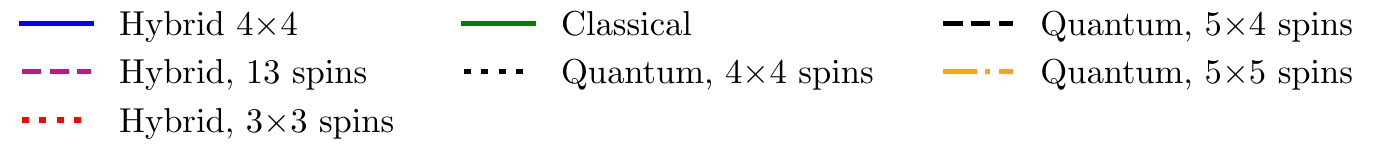}}
 
 \put(5.5, 72)  {\pmark a}
 \put(34.5, 72) {\pmark {a\tprime}}
 
 \put(5.5, 54)  {\pmark b}
 \put(34.5, 54) {\pmark {b\tprime}}
 
 \put(5.5, 36)  {\pmark c}
 \put(34.5, 36) {\pmark {c\tprime}}
 
 \put(5.5, 18)  {\pmark d}
 \put(34.5, 18) {\pmark {d\tprime}}
\end{picture}
\caption{Correlation functions $C_{\alpha} (t)$  for two-dimensional periodic lattices with nearest-neighbour interaction (second set of additional plots). The notations are the same as in Fig.~\ref{onedim} of the main text. For both hybrid and classical simulations, the full lattice size is $9\times9$. The shapes of quantum clusters for hybrid simulations are shown in Fig.~\ref{onedim}(c).
        }
\label{twodim_extra_2}
\end{figure*}

\begin{figure*} \setlength{\unitlength}{0.1in}
\begin{picture}(70,33)
 
 \put(-1, 0) {\includegraphics[width=3.4in]{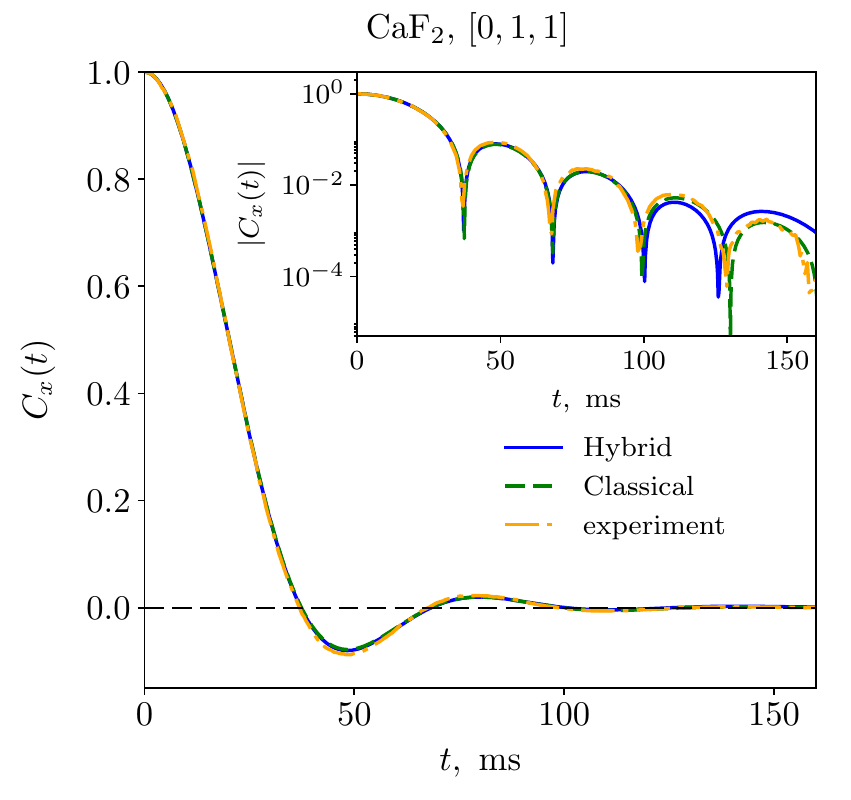}}
 \put(34, 0) {\includegraphics[width=3.4in]{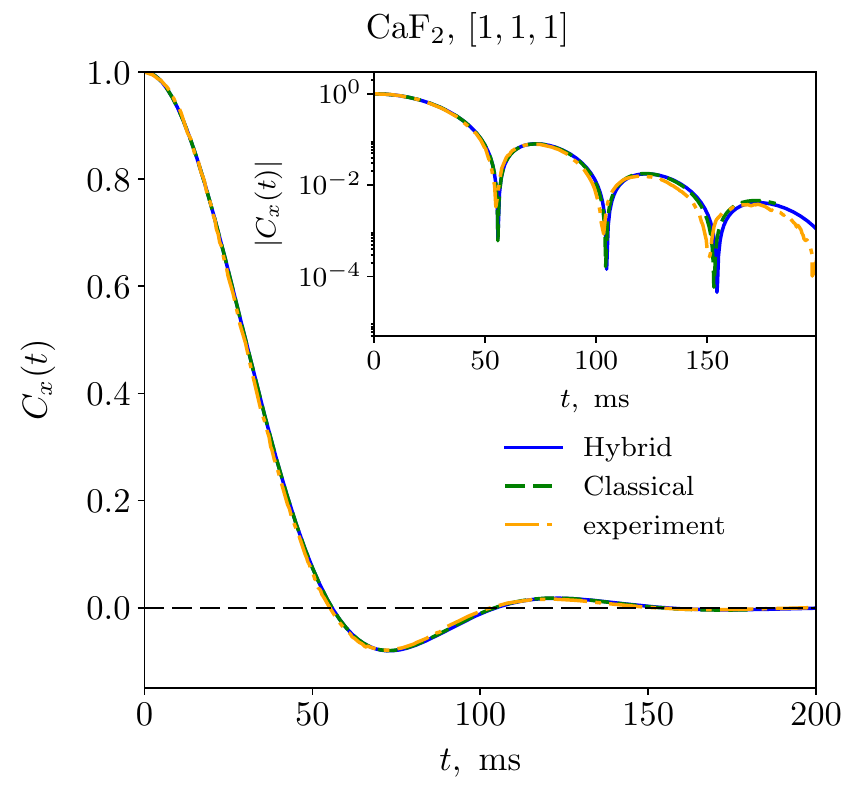}}
 
  \put(-0.5,31) {\pmark a}
 \put(34.5,31) {\pmark b}

\end{picture}
\caption{FIDs in CaF$_2$ for external magnetic field $B_0$ along the following crystal directions: (a) $[011]$; (b) $[111]$. Hybrid and classical simulations are compared with the experimental results of Ref.\cite{Engelsberg-74}. For both hybrid and classical simulations, the full lattice size is $9\times9\times9$. The quantum cluster in hybrid simulations was a chain passing through the entire lattice and oriented along the $x$-axis ([100] crystal direction) in (a)  and  along the main diagonal ([111] crystal direction) in (b). As explained in the main text, these orientations were chosen to maximize the nearest-neighbour couplings within the quantum clusters. The insets contain semi-logarithmic plots of the respective FIDs. 
        }
\label{fid2}
\end{figure*}

We performed additional tests of the hybrid method for one- and two-dimensional lattices  with various sets of nearest-neighbor coupling constants.   The results are presented in Figs.~\ref{onedim_extra}, \ref{twodim_extra_1} and \ref{twodim_extra_2}.
The figures also include plots obtained by direct quantum calculations for different lattice sizes.

Some of these tests are simply complementary to those presented in the main text in the sense that they deal with the same sets of interaction constants but different projections $M_{\alpha}$ of the total spin polarization [Figs.\ref{onedim_extra}(b,b'), \ref{twodim_extra_1}(a,a',b,b')]. 
Other tests are aimed at exploring cases that could be potentially problematic for the hybrid method, including the spin-1/2 $XX$ chain in Figs.~\ref{onedim_extra} (a,a'), which is integrable {\it via} Jordan-Wigner transformation, and the lattices, where the oscillatory and the monotonic long-time modes compete with each other [Figs.~\ref{twodim_extra_2}(b,b',d,d')].
The interaction constants in Figs.~\ref{twodim_extra_1}(a) and \ref{twodim_extra_2}(b) are close to each other, which allows one to follow the evolution of this competition.

We observe that the agreement between the hybrid and the purely quantum results is very good, whenever the quantum results themselves do not exhibit significant finite-size effects [Figs.~\ref{onedim_extra}(a,b), \ref{twodim_extra_1}(a,b) and \ref{twodim_extra_2}(a)]. At the same time, we find that  the competition between different kinds of asymptotic behavior in Figs.~\ref{twodim_extra_1}(a) and \ref{twodim_extra_2}(b,d) is accompanied by larger finite-size effects for the reference plots, which, in turn, makes the tests of the hybrid method not fully conclusive. 

We also tested the performance of the hybrid method for the FID in CaF$_2$ with magnetic field oriented along $[011]$ and $[111]$ crystal directions.  The results are presented on Fig.~\ref{fid2}.
The effective number of interacting neighbours $n_{\text{eff}}$  is significantly larger for the above two directions than for $[001]$\cite{Elsayed-15}. As a result, it was expected and, indeed, observed that the hybrid and the classical results almost coincide.

We finally remark that the hybrid method is supposed to be of most value in those cases, where the direct quantum simulations cannot access the thermodynamic limit for the correlation functions of interest, and, at the same time, the effective number of interacting neighbors $n_{\text{eff}}$ is not large enough to justify purely classical calculations --- for example,  three-dimensional lattices that can be divided into one-dimensional chains with stronger coupling within each chain and weaker coupling between the chains. The performance of the method in such settings should be a subject to future experimental tests.

\section{Statistics behind the plots}

In Table \ref{runtable}, we list the number of computational runs behind the plots presented in both the main text and the supplementary material.

\begin{table*}[t] 
\small
\begin{tabular}{|c|c|c|c|c|}\hline
 Dim. & Coupling constants & Figure  & Plot type& Number of runs\\ \hline
 \multirow{9}{*}{1} &\multirow{3}{*}{$(-0.41,-0.41,0.82)$}     & \multirow{2}{*}{\ref{onedim}(a)}                                & Hybrid, 14 spins         &  $30000$  \\ \cline{4-5}
                    &                                          &                                                                 & Hybrid, 12 spins         &  $13127$  \\ \cline{3-5}
                    &                                          & \ref{onedim}(a\tprime)                                          & Classical                &  $10000$  \\ \cline{2-5}
                    &\multirow{3}{*}{$(0.707,0.707,0.000)$}    & \multirow{2}{*}{\ref{onedim_extra}(a)}                          & Hybrid, 14 spins         &  $30030$  \\ \cline{4-5}
                    &                                          &                                                                 & Hybrid, 12 spins         &  $43078$  \\ \cline{3-5}
                    &                                          & \ref{onedim_extra}(a\tprime)                                    & Classical                &  $10000$  \\ \cline{2-5}
                    &\multirow{3}{*}{$(0.518,0.830,0.207)$}    & \multirow{2}{*}{\ref{onedim}(b,c), \ref{onedim_extra}(b)}       & Hybrid, 16 spins         &  $10860$  \\ \cline{4-5}
                    &                                          &                                                                 & Hybrid, 12 spins         &  $43078$  \\ \cline{3-5}
                    &                                          & \ref{onedim}(b\tprime,c\tprime), \ref{onedim_extra}(b\tprime)   & Classical                &  $10000$  \\ \hline
 \multirow{12}{*}{2}&\multirow{4}{*}{$(-0.41,-0.41,0.82)$}     & \multirow{3}{*}{\ref{twodim}(a)}                                & Hybrid, 4$\times$4 spins &  $41261$  \\ \cline{4-5}
                    &                                          &                                                                 & Hybrid, 13 spins         &  $188000$ \\ \cline{4-5}
                    &                                          &                                                                 & Hybrid, 3$\times$3 spins &  $76203$  \\ \cline{3-5}
                    &                                          & \ref{twodim}(a\tprime)                                          & Classical                &  $16006$  \\ \cline{2-5}
                    &\multirow{3}{*}{$(0.707,0.707,0.00)$}     & \multirow{2}{*}{\ref{twodim_extra_2}(a)}                        & Hybrid, 13 spins         &  $64000$  \\ \cline{4-5}
                    &                                          &                                                                 & Hybrid, 3$\times$3 spins &  $8006$   \\ \cline{3-5}
                    &                                          & \ref{twodim_extra_2}(a\tprime)                                  & Classical                &  $16000$  \\ \cline{2-5}
                    &\multirow{4}{*}{$(0.518,0.830,0.207)$}    & \multirow{3}{*}{\ref{twodim}(b), \ref{twodim_extra_1}(a,b)}     & Hybrid, 4$\times$4 spins &  $5339$   \\ \cline{4-5}
                    &                                          &                                                                 & Hybrid, 13 spins         &  $60000$  \\ \cline{4-5}
                    &                                          &                                                                 & Hybrid, 3$\times$3 spins &  $8006$   \\ \cline{3-5}
                    &                                          & \ref{twodim}(b\tprime), \ref{twodim_extra_1}(a\tprime,b\tprime) & Classical                &  $16006$  \\ \cline{2-5}
                    &\multirow{4}{*}{$(0.400,0.900,0.173)$}    & \multirow{3}{*}{\ref{twodim_extra_2}(b,c,d)}                    & Hybrid, 4$\times$4 spins &  $15699$  \\ \cline{4-5}
                    &                                          &                                                                 & Hybrid, 13 spins         &  $90000$  \\ \cline{4-5}
                    &                                          &                                                                 & Hybrid, 3$\times$3 spins &  $16000$  \\ \cline{3-5}
                    &                                          & \ref{twodim_extra_2}(b\tprime,c\tprime,d\tprime)                & Classical                &  $16000$  \\ \hline
 \multirow{6}{*}{3} & \multirow{2}{*}{CaF$_2$ FID, $[001]$}    & \multirow{2}{*}{\ref{FIDfig}}                                   & Hybrid                   &  $4.3\cdot10^6$ \\ \cline{4-5}
                    &                                          &                                                                 & Classical                &  $4.0\cdot10^5$ \\ \cline{2-5}
                    & \multirow{2}{*}{CaF$_2$ FID, $[011]$}    & \multirow{2}{*}{\ref{fid2}(a)}                                  & Hybrid                   &  $1.4\cdot10^6$ \\ \cline{4-5}
                    &                                          &                                                                 & Classical                &  $4.0\cdot10^5$ \\ \cline{2-5}
                    & \multirow{2}{*}{CaF$_2$ FID, $[111]$}    & \multirow{2}{*}{\ref{fid2}(b)}                                  & Hybrid                   &  $1.1\cdot10^6$ \\ \cline{4-5}
                    &                                          &                                                                 & Classical                &  $4.0\cdot10^5$ \\ \hline

\end{tabular}
\caption{Number of computational runs behind plotted correlation functions. The time length of each run is $10 \, T_0$ or larger, where $T_0$ is the time range where the correlation function is plotted in the respective figure.}
\label{runtable}
\end{table*}


\begin{thebibliography}{38}%
\makeatletter
\providecommand \@ifxundefined [1]{%
 \@ifx{#1\undefined}
}%
\providecommand \@ifnum [1]{%
 \ifnum #1\expandafter \@firstoftwo
 \else \expandafter \@secondoftwo
 \fi
}%
\providecommand \@ifx [1]{%
 \ifx #1\expandafter \@firstoftwo
 \else \expandafter \@secondoftwo
 \fi
}%
\providecommand \natexlab [1]{#1}%
\providecommand \enquote  [1]{``#1''}%
\providecommand \bibnamefont  [1]{#1}%
\providecommand \bibfnamefont [1]{#1}%
\providecommand \citenamefont [1]{#1}%
\providecommand \href@noop [0]{\@secondoftwo}%
\providecommand \href [0]{\begingroup \@sanitize@url \@href}%
\providecommand \@href[1]{\@@startlink{#1}\@@href}%
\providecommand \@@href[1]{\endgroup#1\@@endlink}%
\providecommand \@sanitize@url [0]{\catcode `\\12\catcode `\$12\catcode
  `\&12\catcode `\#12\catcode `\^12\catcode `\_12\catcode `\%12\relax}%
\providecommand \@@startlink[1]{}%
\providecommand \@@endlink[0]{}%
\providecommand \url  [0]{\begingroup\@sanitize@url \@url }%
\providecommand \@url [1]{\endgroup\@href {#1}{\urlprefix }}%
\providecommand \urlprefix  [0]{URL }%
\providecommand \Eprint [0]{\href }%
\providecommand \doibase [0]{http://dx.doi.org/}%
\providecommand \selectlanguage [0]{\@gobble}%
\providecommand \bibinfo  [0]{\@secondoftwo}%
\providecommand \bibfield  [0]{\@secondoftwo}%
\providecommand \translation [1]{[#1]}%
\providecommand \BibitemOpen [0]{}%
\providecommand \bibitemStop [0]{}%
\providecommand \bibitemNoStop [0]{.\EOS\space}%
\providecommand \EOS [0]{\spacefactor3000\relax}%
\providecommand \BibitemShut  [1]{\csname bibitem#1\endcsname}%
\let\auto@bib@innerbib\@empty
\bibitem [{\citenamefont {Lowe}\ and\ \citenamefont {Norberg}(1957)}]{Lowe-57}%
  \BibitemOpen
  \bibfield  {author} {\bibinfo {author} {\bibfnamefont {I.~J.}\ \bibnamefont
  {Lowe}}\ and\ \bibinfo {author} {\bibfnamefont {R.~E.}\ \bibnamefont
  {Norberg}},\ }\href@noop {} {\bibfield  {journal} {\bibinfo  {journal} {Phys.
  Rev.}\ }\textbf {\bibinfo {volume} {107}},\ \bibinfo {pages} {46} (\bibinfo
  {year} {1957})}\BibitemShut {NoStop}%
\bibitem [{\citenamefont {Abragam}(1961)}]{Abragam-61}%
  \BibitemOpen
  \bibfield  {author} {\bibinfo {author} {\bibfnamefont {A.}~\bibnamefont
  {Abragam}},\ }\href@noop {} {\emph {\bibinfo {title} {Principles of Nuclear
  Magnetism}}}\ (\bibinfo  {publisher} {Oxford University Press},\ \bibinfo
  {year} {1961})\BibitemShut {NoStop}%
\bibitem [{\citenamefont {Bloch}(1946)}]{Bloch-46-1}%
  \BibitemOpen
  \bibfield  {author} {\bibinfo {author} {\bibfnamefont {F.}~\bibnamefont
  {Bloch}},\ }\href {\doibase 10.1103/PhysRev.70.460} {\bibfield  {journal}
  {\bibinfo  {journal} {Phys. Rev.}\ }\textbf {\bibinfo {volume} {70}},\
  \bibinfo {pages} {460} (\bibinfo {year} {1946})}\BibitemShut {NoStop}%
\bibitem [{\citenamefont {Fine}\ \emph {et~al.}(2014)\citenamefont {Fine},
  \citenamefont {Elsayed}, \citenamefont {Kropf},\ and\ \citenamefont
  {de~Wijn}}]{Fine-14}%
  \BibitemOpen
  \bibfield  {author} {\bibinfo {author} {\bibfnamefont {B.~V.}\ \bibnamefont
  {Fine}}, \bibinfo {author} {\bibfnamefont {T.~A.}\ \bibnamefont {Elsayed}},
  \bibinfo {author} {\bibfnamefont {C.~M.}\ \bibnamefont {Kropf}}, \ and\
  \bibinfo {author} {\bibfnamefont {A.~S.}\ \bibnamefont {de~Wijn}},\
  }\href@noop {} {\bibfield  {journal} {\bibinfo  {journal} {Phys. Rev. E}\
  }\textbf {\bibinfo {volume} {89}},\ \bibinfo {pages} {012923} (\bibinfo
  {year} {2014})}\BibitemShut {NoStop}%
\bibitem [{\citenamefont {de~Wijn}\ \emph {et~al.}(2012)\citenamefont
  {de~Wijn}, \citenamefont {Hess},\ and\ \citenamefont {Fine}}]{deWijn-12}%
  \BibitemOpen
  \bibfield  {author} {\bibinfo {author} {\bibfnamefont {A.~S.}\ \bibnamefont
  {de~Wijn}}, \bibinfo {author} {\bibfnamefont {B.}~\bibnamefont {Hess}}, \
  and\ \bibinfo {author} {\bibfnamefont {B.~V.}\ \bibnamefont {Fine}},\ }\href
  {\doibase 10.1103/PhysRevLett.109.034101} {\bibfield  {journal} {\bibinfo
  {journal} {Phys. Rev. Lett.}\ }\textbf {\bibinfo {volume} {109}},\ \bibinfo
  {pages} {034101} (\bibinfo {year} {2012})}\BibitemShut {NoStop}%
\bibitem [{\citenamefont {de~Wijn}\ \emph {et~al.}(2013)\citenamefont
  {de~Wijn}, \citenamefont {Hess},\ and\ \citenamefont {Fine}}]{deWijn-13}%
  \BibitemOpen
  \bibfield  {author} {\bibinfo {author} {\bibfnamefont {A.~S.}\ \bibnamefont
  {de~Wijn}}, \bibinfo {author} {\bibfnamefont {B.}~\bibnamefont {Hess}}, \
  and\ \bibinfo {author} {\bibfnamefont {B.~V.}\ \bibnamefont {Fine}},\
  }\href@noop {} {\bibfield  {journal} {\bibinfo  {journal} {Journal of Physics
  A: Mathematical and Theoretical}\ }\textbf {\bibinfo {volume} {46}},\
  \bibinfo {pages} {254012} (\bibinfo {year} {2013})}\BibitemShut {NoStop}%
\bibitem [{\citenamefont {Van~Vleck}(1948)}]{VanVleck-48}%
  \BibitemOpen
  \bibfield  {author} {\bibinfo {author} {\bibfnamefont {J.~H.}\ \bibnamefont
  {Van~Vleck}},\ }\href {\doibase 10.1103/PhysRev.74.1168} {\bibfield
  {journal} {\bibinfo  {journal} {Phys. Rev.}\ }\textbf {\bibinfo {volume}
  {74}},\ \bibinfo {pages} {1168} (\bibinfo {year} {1948})}\BibitemShut
  {NoStop}%
\bibitem [{\citenamefont {Tjon}(1966)}]{Tjon-66}%
  \BibitemOpen
  \bibfield  {author} {\bibinfo {author} {\bibfnamefont {J.~A.}\ \bibnamefont
  {Tjon}},\ }\href {\doibase 10.1103/PhysRev.143.259} {\bibfield  {journal}
  {\bibinfo  {journal} {Phys. Rev.}\ }\textbf {\bibinfo {volume} {143}},\
  \bibinfo {pages} {259} (\bibinfo {year} {1966})}\BibitemShut {NoStop}%
\bibitem [{\citenamefont {Parker}\ and\ \citenamefont
  {Lado}(1973)}]{Parker-73}%
  \BibitemOpen
  \bibfield  {author} {\bibinfo {author} {\bibfnamefont {G.~W.}\ \bibnamefont
  {Parker}}\ and\ \bibinfo {author} {\bibfnamefont {F.}~\bibnamefont {Lado}},\
  }\href@noop {} {\bibfield  {journal} {\bibinfo  {journal} {Phys. Rev. B}\
  }\textbf {\bibinfo {volume} {8}},\ \bibinfo {pages} {3081} (\bibinfo {year}
  {1973})}\BibitemShut {NoStop}%
\bibitem [{\citenamefont {Jensen}\ and\ \citenamefont
  {Platz}(1973)}]{Jensen-73}%
  \BibitemOpen
  \bibfield  {author} {\bibinfo {author} {\bibfnamefont {S.~J.~K.}\
  \bibnamefont {Jensen}}\ and\ \bibinfo {author} {\bibfnamefont
  {O.}~\bibnamefont {Platz}},\ }\href@noop {} {\bibfield  {journal} {\bibinfo
  {journal} {Phys. Rev. B}\ }\textbf {\bibinfo {volume} {7}},\ \bibinfo {pages}
  {31} (\bibinfo {year} {1973})}\BibitemShut {NoStop}%
\bibitem [{\citenamefont {Engelsberg}\ and\ \citenamefont
  {Chao}(1975)}]{Engelsberg-75}%
  \BibitemOpen
  \bibfield  {author} {\bibinfo {author} {\bibfnamefont {M.}~\bibnamefont
  {Engelsberg}}\ and\ \bibinfo {author} {\bibfnamefont {N.-C.}\ \bibnamefont
  {Chao}},\ }\href {\doibase 10.1103/PhysRevB.12.5043} {\bibfield  {journal}
  {\bibinfo  {journal} {Phys. Rev. B}\ }\textbf {\bibinfo {volume} {12}},\
  \bibinfo {pages} {5043} (\bibinfo {year} {1975})}\BibitemShut {NoStop}%
\bibitem [{\citenamefont {Becker}\ \emph {et~al.}(1976)\citenamefont {Becker},
  \citenamefont {Plefka},\ and\ \citenamefont {Sauermann}}]{Becker-76}%
  \BibitemOpen
  \bibfield  {author} {\bibinfo {author} {\bibfnamefont {K.~W.}\ \bibnamefont
  {Becker}}, \bibinfo {author} {\bibfnamefont {T.}~\bibnamefont {Plefka}}, \
  and\ \bibinfo {author} {\bibfnamefont {G.}~\bibnamefont {Sauermann}},\ }\href
  {http://stacks.iop.org/0022-3719/9/i=21/a=023} {\bibfield  {journal}
  {\bibinfo  {journal} {Journal of Physics C: Solid State Physics}\ }\textbf
  {\bibinfo {volume} {9}},\ \bibinfo {pages} {4041} (\bibinfo {year}
  {1976})}\BibitemShut {NoStop}%
\bibitem [{\citenamefont {Shakhmuratov}(1991)}]{Shakhmuratov-91}%
  \BibitemOpen
  \bibfield  {author} {\bibinfo {author} {\bibfnamefont {R.~N.}\ \bibnamefont
  {Shakhmuratov}},\ }\href {http://stacks.iop.org/0953-8984/3/i=44/a=013}
  {\bibfield  {journal} {\bibinfo  {journal} {Journal of Physics: Condensed
  Matter}\ }\textbf {\bibinfo {volume} {3}},\ \bibinfo {pages} {8683} (\bibinfo
  {year} {1991})}\BibitemShut {NoStop}%
\bibitem [{\citenamefont {Lundin}(1992)}]{Lundin-92}%
  \BibitemOpen
  \bibfield  {author} {\bibinfo {author} {\bibfnamefont {A.~A.}\ \bibnamefont
  {Lundin}},\ }\href@noop {} {\bibfield  {journal} {\bibinfo  {journal} {Sov.
  Phys. JETP}\ }\textbf {\bibinfo {volume} {102}},\ \bibinfo {pages} {352}
  (\bibinfo {year} {1992})}\BibitemShut {NoStop}%
\bibitem [{\citenamefont {Jensen}(1995)}]{Jensen-95}%
  \BibitemOpen
  \bibfield  {author} {\bibinfo {author} {\bibfnamefont {J.}~\bibnamefont
  {Jensen}},\ }\href {\doibase 10.1103/PhysRevB.52.9611} {\bibfield  {journal}
  {\bibinfo  {journal} {Phys. Rev. B}\ }\textbf {\bibinfo {volume} {52}},\
  \bibinfo {pages} {9611} (\bibinfo {year} {1995})}\BibitemShut {NoStop}%
\bibitem [{\citenamefont {Fine}(1997)}]{Fine-97}%
  \BibitemOpen
  \bibfield  {author} {\bibinfo {author} {\bibfnamefont {B.~V.}\ \bibnamefont
  {Fine}},\ }\href@noop {} {\bibfield  {journal} {\bibinfo  {journal} {Phys.
  Rev. Lett.}\ }\textbf {\bibinfo {volume} {79}},\ \bibinfo {pages} {4673}
  (\bibinfo {year} {1997})}\BibitemShut {NoStop}%
\bibitem [{\citenamefont {Zhang}\ \emph {et~al.}(2007)\citenamefont {Zhang},
  \citenamefont {Konstantinidis}, \citenamefont {{Al-Hassanieh}},\ and\
  \citenamefont {Dobrovitski}}]{Zhang-07}%
  \BibitemOpen
  \bibfield  {author} {\bibinfo {author} {\bibfnamefont {W.}~\bibnamefont
  {Zhang}}, \bibinfo {author} {\bibfnamefont {N.}~\bibnamefont
  {Konstantinidis}}, \bibinfo {author} {\bibfnamefont {K.~A.}\ \bibnamefont
  {{Al-Hassanieh}}}, \ and\ \bibinfo {author} {\bibfnamefont {V.~V.}\
  \bibnamefont {Dobrovitski}},\ }\href@noop {} {\bibfield  {journal} {\bibinfo
  {journal} {J. Phys. Condens. Matter}\ }\textbf {\bibinfo {volume} {19}},\
  \bibinfo {pages} {083202} (\bibinfo {year} {2007})}\BibitemShut {NoStop}%
\bibitem [{\citenamefont {Savostyanov}\ \emph {et~al.}(2014)\citenamefont
  {Savostyanov}, \citenamefont {Dolgov}, \citenamefont {Werner},\ and\
  \citenamefont {Kuprov}}]{Savostyanov-14}%
  \BibitemOpen
  \bibfield  {author} {\bibinfo {author} {\bibfnamefont {D.~V.}\ \bibnamefont
  {Savostyanov}}, \bibinfo {author} {\bibfnamefont {S.~V.}\ \bibnamefont
  {Dolgov}}, \bibinfo {author} {\bibfnamefont {J.~M.}\ \bibnamefont {Werner}},
  \ and\ \bibinfo {author} {\bibfnamefont {I.}~\bibnamefont {Kuprov}},\ }\href
  {\doibase 10.1103/PhysRevB.90.085139} {\bibfield  {journal} {\bibinfo
  {journal} {Phys. Rev. B}\ }\textbf {\bibinfo {volume} {90}},\ \bibinfo
  {pages} {085139} (\bibinfo {year} {2014})}\BibitemShut {NoStop}%
\bibitem [{\citenamefont {Elsayed}\ and\ \citenamefont
  {Fine}(2015)}]{Elsayed-15}%
  \BibitemOpen
  \bibfield  {author} {\bibinfo {author} {\bibfnamefont {T.~A.}\ \bibnamefont
  {Elsayed}}\ and\ \bibinfo {author} {\bibfnamefont {B.~V.}\ \bibnamefont
  {Fine}},\ }\href@noop {} {\bibfield  {journal} {\bibinfo  {journal} {Phys.
  Rev. B}\ }\textbf {\bibinfo {volume} {91}},\ \bibinfo {pages} {094424}
  (\bibinfo {year} {2015})}\BibitemShut {NoStop}%
\bibitem [{\citenamefont {Engelsberg}\ and\ \citenamefont
  {Lowe}(1974)}]{Engelsberg-74}%
  \BibitemOpen
  \bibfield  {author} {\bibinfo {author} {\bibfnamefont {M.}~\bibnamefont
  {Engelsberg}}\ and\ \bibinfo {author} {\bibfnamefont {I.~J.}\ \bibnamefont
  {Lowe}},\ }\href@noop {} {\bibfield  {journal} {\bibinfo  {journal} {Phys.
  Rev. B}\ }\textbf {\bibinfo {volume} {10}},\ \bibinfo {pages} {822} (\bibinfo
  {year} {1974})}\BibitemShut {NoStop}%
\bibitem [{\citenamefont {Coish}\ and\ \citenamefont {Loss}(2005)}]{Coish-05}%
  \BibitemOpen
  \bibfield  {author} {\bibinfo {author} {\bibfnamefont {W.~A.}\ \bibnamefont
  {Coish}}\ and\ \bibinfo {author} {\bibfnamefont {D.}~\bibnamefont {Loss}},\
  }\href {\doibase 10.1103/PhysRevB.72.125337} {\bibfield  {journal} {\bibinfo
  {journal} {Phys. Rev. B}\ }\textbf {\bibinfo {volume} {72}},\ \bibinfo
  {pages} {125337} (\bibinfo {year} {2005})}\BibitemShut {NoStop}%
\bibitem [{\citenamefont {Al-Hassanieh}\ \emph {et~al.}(2006)\citenamefont
  {Al-Hassanieh}, \citenamefont {Dobrovitski}, \citenamefont {Dagotto},\ and\
  \citenamefont {Harmon}}]{Dobrovitski-06}%
  \BibitemOpen
  \bibfield  {author} {\bibinfo {author} {\bibfnamefont {K.~A.}\ \bibnamefont
  {Al-Hassanieh}}, \bibinfo {author} {\bibfnamefont {V.~V.}\ \bibnamefont
  {Dobrovitski}}, \bibinfo {author} {\bibfnamefont {E.}~\bibnamefont
  {Dagotto}}, \ and\ \bibinfo {author} {\bibfnamefont {B.~N.}\ \bibnamefont
  {Harmon}},\ }\href@noop {} {\bibfield  {journal} {\bibinfo  {journal} {Phys.
  Rev. Lett.}\ }\textbf {\bibinfo {volume} {97}},\ \bibinfo {pages} {037204}
  (\bibinfo {year} {2006})}\BibitemShut {NoStop}%
\bibitem [{\citenamefont {Liu}\ \emph {et~al.}(2007)\citenamefont {Liu},
  \citenamefont {Yao},\ and\ \citenamefont {Sham}}]{Liu-07}%
  \BibitemOpen
  \bibfield  {author} {\bibinfo {author} {\bibfnamefont {R.-B.}\ \bibnamefont
  {Liu}}, \bibinfo {author} {\bibfnamefont {W.}~\bibnamefont {Yao}}, \ and\
  \bibinfo {author} {\bibfnamefont {L.~J.}\ \bibnamefont {Sham}},\ }\href
  {http://stacks.iop.org/1367-2630/9/i=7/a=226} {\bibfield  {journal} {\bibinfo
   {journal} {New Journal of Physics}\ }\textbf {\bibinfo {volume} {9}},\
  \bibinfo {pages} {226} (\bibinfo {year} {2007})}\BibitemShut {NoStop}%
\bibitem [{\citenamefont {Stanwix}\ \emph {et~al.}(2010)\citenamefont
  {Stanwix}, \citenamefont {Pham}, \citenamefont {Maze}, \citenamefont
  {Le~Sage}, \citenamefont {Yeung}, \citenamefont {Cappellaro}, \citenamefont
  {Hemmer}, \citenamefont {Yacoby}, \citenamefont {Lukin},\ and\ \citenamefont
  {Walsworth}}]{Stanwix-10}%
  \BibitemOpen
  \bibfield  {author} {\bibinfo {author} {\bibfnamefont {P.~L.}\ \bibnamefont
  {Stanwix}}, \bibinfo {author} {\bibfnamefont {L.~M.}\ \bibnamefont {Pham}},
  \bibinfo {author} {\bibfnamefont {J.~R.}\ \bibnamefont {Maze}}, \bibinfo
  {author} {\bibfnamefont {D.}~\bibnamefont {Le~Sage}}, \bibinfo {author}
  {\bibfnamefont {T.~K.}\ \bibnamefont {Yeung}}, \bibinfo {author}
  {\bibfnamefont {P.}~\bibnamefont {Cappellaro}}, \bibinfo {author}
  {\bibfnamefont {P.~R.}\ \bibnamefont {Hemmer}}, \bibinfo {author}
  {\bibfnamefont {A.}~\bibnamefont {Yacoby}}, \bibinfo {author} {\bibfnamefont
  {M.~D.}\ \bibnamefont {Lukin}}, \ and\ \bibinfo {author} {\bibfnamefont
  {R.~L.}\ \bibnamefont {Walsworth}},\ }\href {\doibase
  10.1103/PhysRevB.82.201201} {\bibfield  {journal} {\bibinfo  {journal} {Phys.
  Rev. B}\ }\textbf {\bibinfo {volume} {82}},\ \bibinfo {pages} {201201}
  (\bibinfo {year} {2010})}\BibitemShut {NoStop}%
\bibitem [{\citenamefont {Balcar}\ and\ \citenamefont
  {Lovesey}(1989)}]{Balcar-81}%
  \BibitemOpen
  \bibfield  {author} {\bibinfo {author} {\bibfnamefont {E.}~\bibnamefont
  {Balcar}}\ and\ \bibinfo {author} {\bibfnamefont {S.~W.}\ \bibnamefont
  {Lovesey}},\ }\href@noop {} {\emph {\bibinfo {title} {Theory of Magnetic
  Neutron and Photon Scattering}}}\ (\bibinfo  {publisher} {Clarendon Press},\
  \bibinfo {year} {1989})\BibitemShut {NoStop}%
\bibitem [{\citenamefont {Gemmer}\ \emph {et~al.}(2004)\citenamefont {Gemmer},
  \citenamefont {Michel},\ and\ \citenamefont {Mahler}}]{Gemmer-04}%
  \BibitemOpen
  \bibfield  {author} {\bibinfo {author} {\bibfnamefont {J.}~\bibnamefont
  {Gemmer}}, \bibinfo {author} {\bibfnamefont {M.}~\bibnamefont {Michel}}, \
  and\ \bibinfo {author} {\bibfnamefont {G.}~\bibnamefont {Mahler}},\
  }\href@noop {} {\emph {\bibinfo {title} {Quantum Thermodynamics}}}\ (\bibinfo
   {publisher} {Springer-Verlag Berlin Heidelberg},\ \bibinfo {year}
  {2004})\BibitemShut {NoStop}%
\bibitem [{\citenamefont {\'Alvarez}\ \emph {et~al.}(2008)\citenamefont
  {\'Alvarez}, \citenamefont {Danieli}, \citenamefont {Levstein},\ and\
  \citenamefont {Pastawski}}]{Pastawski-08}%
  \BibitemOpen
  \bibfield  {author} {\bibinfo {author} {\bibfnamefont {G.~A.}\ \bibnamefont
  {\'Alvarez}}, \bibinfo {author} {\bibfnamefont {E.~P.}\ \bibnamefont
  {Danieli}}, \bibinfo {author} {\bibfnamefont {P.~R.}\ \bibnamefont
  {Levstein}}, \ and\ \bibinfo {author} {\bibfnamefont {H.~M.}\ \bibnamefont
  {Pastawski}},\ }\href {\doibase 10.1103/PhysRevLett.101.120503} {\bibfield
  {journal} {\bibinfo  {journal} {Phys. Rev. Lett.}\ }\textbf {\bibinfo
  {volume} {101}},\ \bibinfo {pages} {120503} (\bibinfo {year}
  {2008})}\BibitemShut {NoStop}%
\bibitem [{\citenamefont {Elsayed}\ and\ \citenamefont
  {Fine}(2013)}]{Elsayed-13}%
  \BibitemOpen
  \bibfield  {author} {\bibinfo {author} {\bibfnamefont {T.~A.}\ \bibnamefont
  {Elsayed}}\ and\ \bibinfo {author} {\bibfnamefont {B.~V.}\ \bibnamefont
  {Fine}},\ }\href@noop {} {\bibfield  {journal} {\bibinfo  {journal} {Phys.
  Rev. Lett.}\ }\textbf {\bibinfo {volume} {110}},\ \bibinfo {pages} {070404}
  (\bibinfo {year} {2013})}\BibitemShut {NoStop}%
\bibitem [{\citenamefont {Morgan}\ \emph {et~al.}(2008)\citenamefont {Morgan},
  \citenamefont {Fine},\ and\ \citenamefont {Saam}}]{Morgan-08}%
  \BibitemOpen
  \bibfield  {author} {\bibinfo {author} {\bibfnamefont {S.~W.}\ \bibnamefont
  {Morgan}}, \bibinfo {author} {\bibfnamefont {B.~V.}\ \bibnamefont {Fine}}, \
  and\ \bibinfo {author} {\bibfnamefont {B.}~\bibnamefont {Saam}},\ }\href@noop
  {} {\bibfield  {journal} {\bibinfo  {journal} {Phys. Rev. Lett.}\ }\textbf
  {\bibinfo {volume} {101}},\ \bibinfo {pages} {067601} (\bibinfo {year}
  {2008})}\BibitemShut {NoStop}%
\bibitem [{\citenamefont {Sorte}\ \emph {et~al.}(2011)\citenamefont {Sorte},
  \citenamefont {Fine},\ and\ \citenamefont {Saam}}]{Sorte-11}%
  \BibitemOpen
  \bibfield  {author} {\bibinfo {author} {\bibfnamefont {E.~G.}\ \bibnamefont
  {Sorte}}, \bibinfo {author} {\bibfnamefont {B.~V.}\ \bibnamefont {Fine}}, \
  and\ \bibinfo {author} {\bibfnamefont {B.}~\bibnamefont {Saam}},\ }\href@noop
  {} {\bibfield  {journal} {\bibinfo  {journal} {Phys. Rev. B}\ }\textbf
  {\bibinfo {volume} {83}},\ \bibinfo {pages} {064302} (\bibinfo {year}
  {2011})}\BibitemShut {NoStop}%
\bibitem [{\citenamefont {Meier}\ \emph {et~al.}(2012)\citenamefont {Meier},
  \citenamefont {Kohlrautz},\ and\ \citenamefont {Haase}}]{Meier-12}%
  \BibitemOpen
  \bibfield  {author} {\bibinfo {author} {\bibfnamefont {B.}~\bibnamefont
  {Meier}}, \bibinfo {author} {\bibfnamefont {J.}~\bibnamefont {Kohlrautz}}, \
  and\ \bibinfo {author} {\bibfnamefont {J.}~\bibnamefont {Haase}},\
  }\href@noop {} {\bibfield  {journal} {\bibinfo  {journal} {Phys. Rev. Lett.}\
  }\textbf {\bibinfo {volume} {108}},\ \bibinfo {pages} {177602} (\bibinfo
  {year} {2012})}\BibitemShut {NoStop}%
\bibitem [{\citenamefont {Fabricius}\ \emph {et~al.}(1997)\citenamefont
  {Fabricius}, \citenamefont {L{\"{o}}w},\ and\ \citenamefont
  {Stolze}}]{Fabricius-97}%
  \BibitemOpen
  \bibfield  {author} {\bibinfo {author} {\bibfnamefont {K.}~\bibnamefont
  {Fabricius}}, \bibinfo {author} {\bibfnamefont {U.}~\bibnamefont
  {L{\"{o}}w}}, \ and\ \bibinfo {author} {\bibfnamefont {J.}~\bibnamefont
  {Stolze}},\ }\href@noop {} {\bibfield  {journal} {\bibinfo  {journal} {Phys.
  Rev. B}\ }\textbf {\bibinfo {volume} {55}},\ \bibinfo {pages} {5833}
  (\bibinfo {year} {1997})}\BibitemShut {NoStop}%
\bibitem [{\citenamefont {Fine}(2003)}]{Fine-03}%
  \BibitemOpen
  \bibfield  {author} {\bibinfo {author} {\bibfnamefont {B.~V.}\ \bibnamefont
  {Fine}},\ }\href@noop {} {\bibfield  {journal} {\bibinfo  {journal} {J. Stat.
  Phys.}\ }\textbf {\bibinfo {volume} {112}},\ \bibinfo {pages} {319} (\bibinfo
  {year} {2003})}\BibitemShut {NoStop}%
\bibitem [{\citenamefont {Borckmans}\ and\ \citenamefont
  {Walgraef}(1968)}]{Borckmans-68}%
  \BibitemOpen
  \bibfield  {author} {\bibinfo {author} {\bibfnamefont {P.}~\bibnamefont
  {Borckmans}}\ and\ \bibinfo {author} {\bibfnamefont {D.}~\bibnamefont
  {Walgraef}},\ }\href {\doibase 10.1103/PhysRev.167.282} {\bibfield  {journal}
  {\bibinfo  {journal} {Phys. Rev.}\ }\textbf {\bibinfo {volume} {167}},\
  \bibinfo {pages} {282} (\bibinfo {year} {1968})}\BibitemShut {NoStop}%
\bibitem [{\citenamefont {Fine}(2004)}]{Fine-04}%
  \BibitemOpen
  \bibfield  {author} {\bibinfo {author} {\bibfnamefont {B.~V.}\ \bibnamefont
  {Fine}},\ }\href@noop {} {\bibfield  {journal} {\bibinfo  {journal} {Int. J.
  Mod. Phys. B}\ }\textbf {\bibinfo {volume} {18}},\ \bibinfo {pages} {1119}
  (\bibinfo {year} {2004})}\BibitemShut {NoStop}%
\bibitem [{\citenamefont {Fine}(2005)}]{Fine-05}%
  \BibitemOpen
  \bibfield  {author} {\bibinfo {author} {\bibfnamefont {B.~V.}\ \bibnamefont
  {Fine}},\ }\href@noop {} {\bibfield  {journal} {\bibinfo  {journal} {Phys.
  Rev. Lett.}\ }\textbf {\bibinfo {volume} {94}},\ \bibinfo {pages} {247601}
  (\bibinfo {year} {2005})}\BibitemShut {NoStop}%
\bibitem [{\citenamefont {Tang}\ and\ \citenamefont {Waugh}(1992)}]{Tang-92}%
  \BibitemOpen
  \bibfield  {author} {\bibinfo {author} {\bibfnamefont {C.}~\bibnamefont
  {Tang}}\ and\ \bibinfo {author} {\bibfnamefont {J.~S.}\ \bibnamefont
  {Waugh}},\ }\href@noop {} {\bibfield  {journal} {\bibinfo  {journal} {Phys.
  Rev. B}\ }\textbf {\bibinfo {volume} {45}},\ \bibinfo {pages} {748} (\bibinfo
  {year} {1992})}\BibitemShut {NoStop}%
\bibitem [{\citenamefont {Lundin}\ and\ \citenamefont
  {Zobov}(1977)}]{Lundin-77}%
  \BibitemOpen
  \bibfield  {author} {\bibinfo {author} {\bibfnamefont {A.~A.}\ \bibnamefont
  {Lundin}}\ and\ \bibinfo {author} {\bibfnamefont {V.~E.}\ \bibnamefont
  {Zobov}},\ }\href@noop {} {\bibfield  {journal} {\bibinfo  {journal} {J.
  Magn. Reson.}\ }\textbf {\bibinfo {volume} {26}},\ \bibinfo {pages} {229}
  (\bibinfo {year} {1977})}\BibitemShut {NoStop}%
\end{thebibliography}
\end{document}